\documentclass[prd,aps,twocolumn,nofootinbib,preprintnumbers,superscriptaddress,preprintnumbers,balancelastpage,longbibliography]{revtex4-2}
\pdfoutput=1
\usepackage{amsmath,mathtools,physics,xfrac}
\usepackage{graphicx}
\usepackage{afterpage}
\usepackage{float}
\usepackage{subfigure}
\usepackage{rotating}
\usepackage{multirow}
\usepackage{tabularx}
\usepackage{booktabs}
\usepackage{fancyhdr}
\usepackage[utf8]{inputenc}
\usepackage{theorem}
\usepackage{moreverb}
\usepackage{euscript}
\usepackage{psfrag}
\usepackage{slashed}
\usepackage{mathtools}
\usepackage{makecell}
\usepackage{adjustbox}
\usepackage{dcolumn}
\usepackage{bm}
\usepackage[dvipsnames]{xcolor}
\usepackage{graphics}
\usepackage{hyperref}
\usepackage[shortlabels]{enumitem}

\hypersetup{
     colorlinks   = true,
     citecolor    = purple,
     urlcolor     = purple,
     linkcolor    = purple
}

\begin{document}

\preprint{UCI-HEP-TR-2024-10}

\title{X-Ray Constraints on Dark Photon Tridents}

\author{Tim Linden}
\thanks{\href{mailto:linden@fysik.su.se}{linden@fysik.su.se}, \href{https://orcid.org/0000-0001-9888-0971}{ORCID: 0000-0001-9888-0971}}
\affiliation{Stockholm University and The Oskar Klein Centre for Cosmoparticle Physics,  Alba Nova, 10691 Stockholm, Sweden}

\author{Thong T.Q. Nguyen}
\thanks{\href{mailto:thong.nguyen@fysik.su.se}{thong.nguyen@fysik.su.se}, \href{https://orcid.org/0000-0002-8460-0219}{ORCID: 0000-0002-8460-0219}}
\affiliation{Stockholm University and The Oskar Klein Centre for Cosmoparticle Physics,  Alba Nova, 10691 Stockholm, Sweden}

\author{Tim M.P. Tait}
\thanks{\href{mailto:ttait@uci.edu}{ttait@uci.edu}, \href{https://orcid.org/0000-0003-3002-6909}{ORCID: 0000-0003-3002-6909}}
\affiliation{Department of Physics and Astronomy, University of California, Irvine, CA 92697 USA}

\begin{abstract}
\noindent Dark photons that are sufficiently light and/or weakly-interacting represent a compelling vision of dark matter. Dark photon decay into three photons, which we call the dark photon trident, can be the dominant channel when the dark photon mass falls below the electron pair threshold and can produce a significant flux of x-rays. We use 16 years of data from INTEGRAL/SPI to constrain sub-MeV dark photon decay, producing new worlds-best constraints on the kinetic mixing parameter for dark photon masses between 90 keV and 1022 keV, and comment on the potential for future x-ray observatories to discover the trident decay process.
\end{abstract}

\maketitle

\noindent\textbf{\emph{Introduction.}} --- 
The dark photon is a hypothetical massive spin-1 particle that exists in many extensions of the Standard Model (SM) incorporating an extra $U(1)$ gauge symmetry whose corresponding gauge boson interacts with the SM fermions via kinetic mixing with the SM photon~\cite{Holdom:1985ag,Fayet:1980ad,Fayet:1990wx,Fabbrichesi:2020wbt, Rizzo:2018ntg, Pospelov:2008zw, Nelson:2011sf}.  It is typically categorized as a Weakly Interacting Slim Particle (WISP)~\cite{Caputo:2021eaa}.  Dark photons are the target of a rich array of both high-energy and astrophysical searches for Beyond Standard Model (BSM) particles~\cite{Redondo:2008ec, Arias:2012az, Bloch:2016sjj, XENON:2019gfn, XENON:2020rca, XENON:2021qze, An:2020bxd, Fischbach:1994ir, Zechlin:2008tj, Bi:2020ths, Tran:2023lzv, Li:2023vpv, Wadekar:2019mpc, Dolan:2023cjs, Dubovsky:2015cca, Yan:2023kdg, Hong:2020bxo, Vinyoles:2015aba}.

Provided its mass and kinetic mixing are chosen such that it is stable on cosmological time scales, a dark photon can play the role of the dark matter~\cite{Arias:2012az, Bertone:2004pz,Bertone:2016nfn, Bertone:2018krk, Nelson:2011sf, Servant:2002aq}. However, non-zero kinetic mixing implies that dark photons eventually decay into potentially observable SM particles. For a given mass, the end products of decay are determined by the kinematically accessible SM particles. The subsequent interactions of these particles can produce a distribution of bright photon signals that can be detected across the electromagnetic spectrum.

The situation is much simpler, however, if the dark photon mass falls below twice the electron mass. In this case, only photon and neutrino final states are kinematically accessible. Since the two photon final state is forbidden by the Landau-Yang theorem \cite{Landau:1948kw,Yang:1950rg} and decays to neutrinos are highly suppressed, the dominant channel is the three photon final state through a process which we term the `dark photon trident'. Decays into photon tridents produce a unique spectrum that is broader than the line spectrum from two photon states~\cite{Pospelov:2007mp}. While a combination of astrophysical and direct-detection constraints rule out large kinetic couplings through much of the sub-MeV mass range~\cite{Caputo:2021eaa}, there are notable sensitivity gaps in the x-ray band.

The primary difficulty in constraining dark photon decays stems from the extremely diffuse morphology of the resulting x-ray signal. For standard Navarro-Frenk White (NFW) profiles, the most sensitive decay searches are those which observe as much of the sky as possible. This makes investigations with high-resolution telescopes such as $Chandra$ or XMM-Newton difficult. However, Wide-field-of-view telescopes, such as INTEGRAL, are highly effective for dark matter decay searches.

\begin{figure}[!t]
\centering
\includegraphics[width=\columnwidth]{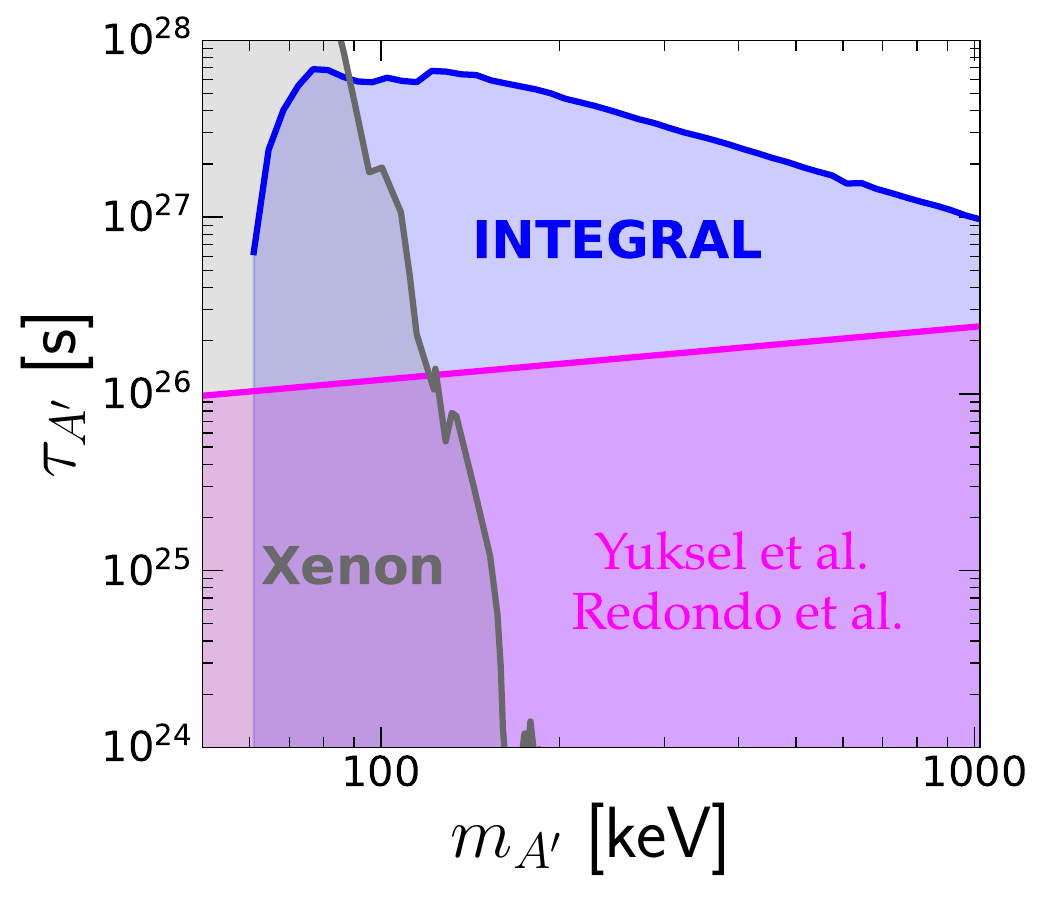}
\caption{95\% C.L. \textcolor{blue}{INTEGRAL} upper limits (\textcolor{blue}{blue line}) on the dark photon dark matter lifetime based on the three photon decay channel. \textcolor{magenta}{Previous constraints} based on the diffuse X-ray background in \textcolor{magenta}{magenta} from Ref.~\cite{Yuksel:2007dr} (model independent) and from Refs.~{\cite{Redondo:2008ec, Arias:2012az}} are shown for the same dark photon trident. Direct detection constraints from \textcolor{gray}{Xenon} are shown in \textcolor{gray}{gray}~\cite{Bloch:2016sjj, XENON:2019gfn, XENON:2020rca, XENON:2021qze, An:2020bxd}. }
\label{fig:lifetime}
\end{figure}

In this {\it Letter}, we use 16-years of x-ray data from INTEGRAL/SPI to strongly constrain dark matter decays via photon tridents in the regime where the dark photon mass is below twice the electron mass. Figure~\ref{fig:lifetime} shows our result, which improves current constraints on the dark photon lifetime by more than an order of magnitude between 61--1022~keV. Compared to previous work, our limits are significantly strengthened by the inclusion of the full one-loop amplitude for the dark photon decay, in contrast to the leading Euler-Heisenberg approximation used in previous analyses~\cite{Redondo:2008ec, Arias:2012az}. Our results strongly motivate further dark photon searches with next-generation x-ray telescopes. \\


\vspace{3pt}
\noindent\textbf{\emph{Dark photon trident.}} --- The Lagrangian describing a massive dark photon $m_{A^{\prime}}$ and its interactions with SM particles is given by:
\begin{equation}
    \mathcal{L}\supset -\frac{1}{4}F_{\mu\nu}^{\prime}F^{\prime\mu\nu}-\frac{1}{2}m_{A^{\prime}}^{2}A_{\mu}^{\prime}A^{\prime\mu}-\frac{\epsilon}{2}F^{\prime}_{\mu\nu}F^{\mu\nu},
    \label{eq:lagrange}
\end{equation}
where the field strength tensors are $F^{\prime}_{\mu\nu}=\partial_{\mu}A^{\prime}_{\nu}-\partial_{\nu}A^{\prime}_{\mu}$, etc. For non-zero values of $\epsilon$, the dark photon kinetically mixes with the SM photon, allowing it to decay into any SM particles that also interact with the photon.



For dark photon masses below twice the electron mass, the dominant decay channels include three-photon final states, which we show in Figure~\ref{fig:diagram} and call the dark photon trident. The only other kinematically allowed final states are SM neutrinos, which result from mixing between the dark photon and the SM $Z$ boson~(see Refs. \cite{Nguyen:2022zwb, Nguyen:2025ygc, Nguyen:2023ugx} for discussion of the indirect searches). However, these are strongly suppressed by the large $Z$ mass, and are ignored throughout the remainder of this work. Previous work~\cite{Pospelov:2008zw, Redondo:2008ec} computed the photon trident rate using the leading dimension-8 term in an Effective Field Theory (EFT) obtained (for $m_A \ll 2 m_e$) by integrating out heavy SM fermion fields at one-loop (see Fig.~\ref{fig:diagram}):
%
 %
\begin{equation}
\begin{split}
    \mathcal{L}^{\rm EH}_{A^{\prime}}=\frac{\epsilon \alpha_{\rm em}}{45 m_{e}^{4}}\Big{(}14F^{\prime}_{\mu\nu}F^{\nu\lambda}&F_{\lambda\rho}F^{\rho\mu}\\
    &-5 F^{\prime}_{\mu\nu}F^{\mu\nu}F_{\alpha\beta}F^{\alpha\beta}\Big{)},
\end{split}
\label{eq:EHlagrangian}
\end{equation}
which results in an expression for the decay width in the leading EFT (Euler-Heisenberg) limit~\cite{Pospelov:2007mp}:
\begin{equation}
    \begin{split}
        \Gamma_{\rm EH}&=\frac{17 \epsilon^{2}\alpha^{4}_{\rm em}}{11664000\pi^{3}}\times \frac{m^{9}_{A^{\prime}}}{m^{8}_{e}}\\
        &\simeq 1 {\rm s}^{-1}\times\Big{(}\frac{\epsilon}{0.003}\Big{)}^{2}\times\Big{(}\frac{m_{A^{\prime}}}{m_{e}}\Big{)}^{9}.
    \end{split}
    \label{eq:GEH}
\end{equation}
However, the leading EFT approximation breaks down for $m_{A^{\prime}} \sim m_e$. The full decay width can be related to the Euler-Heisenberg approximation \cite{McDermott:2017qcg}:
\begin{equation}
    \Gamma_{A^{\prime}\to 3\gamma}=\Gamma_{\rm EH}\times f_{\rm loop}(m_{A^{\prime}}),
\end{equation}
where $f_{\rm loop}(m_{A^{\prime}})$ is an enhancement factor that depends on the dark 
photon mass, computed in Ref.~\cite{McDermott:2017qcg}

Throughout this work, we assume the dark photon makes up the entirety of the dark matter in the Universe. 
It can be produced via the misalignment mechanism and/or non-thermal production during inflation, when the DM mass arises through the St{\"u}eckeblberg mechanism~\cite{Arias:2012az, Nelson:2011sf}. 
The photon flux from dark photon decays is:
\begin{equation}
    \frac{{\rm d}\Phi_{\gamma}}{{\rm d}E_{\gamma}}=\frac{\Gamma}{4\pi m_{\rm DM}}\times\frac{{\rm d}N_{\gamma}}{{\rm d}E_{\gamma}}\times D,
    \label{eq:flux}
\end{equation}
where $\Gamma\equiv \Gamma_{A^{\prime}\to 3\gamma}$ and $m_{\rm DM}\equiv m_{A^{\prime}}$. The energy spectrum for dark photon trident decay is given by~\cite{Pospelov:2007mp, Linden:2024uph}:
\begin{equation}
    \frac{{\rm d}N_{\gamma}}{{\rm d}E_{\gamma}}=\frac{2x^{3}}{17 m_{A^{\prime}}}(1715-3105x+\frac{2919}{2}x^{2}),
    \label{eq:spectrum}
\end{equation}
where $x=2E_{\gamma}/m_{A^{\prime}}$ runs between 0 and 1. The $D-$factor in Eq.~\ref{eq:flux} depends on the dark matter density profile of the Milky Way, integrated over the line of sight $s$:
\begin{equation}
    D=\int {\rm d}\Omega\int\ {\rm d}s \rho\Big{(}r( s, l, b)\Big{)}.
    \label{eq:D}
\end{equation}
where $l$ and $b$ are Galactic longitude and latitude, and the differential solid angle ${\rm d\Omega}=\cos{l}{\rm d}l{\rm d}b$. We assume a Navarro-Frenk-White (NFW) dark matter density profile with $r_s$~=~9.98~kpc and $\rho_{\odot}$~=~0.42~GeV/cm$^{3}$~\cite{Navarro:1996gj}. For the region of interest (ROI) in the Galactic halo with longitude $ |l| \leq 47.5^{\circ}$ and latitude $|b|\leq 47.5^{\circ}$, we obtain \mbox{$D=1.3\times 10^{23}$~GeV/cm$^{2}$~\cite{Karukes:2019jwa}.}

\begin{figure}[!t]
\centering
\includegraphics[width=\columnwidth]{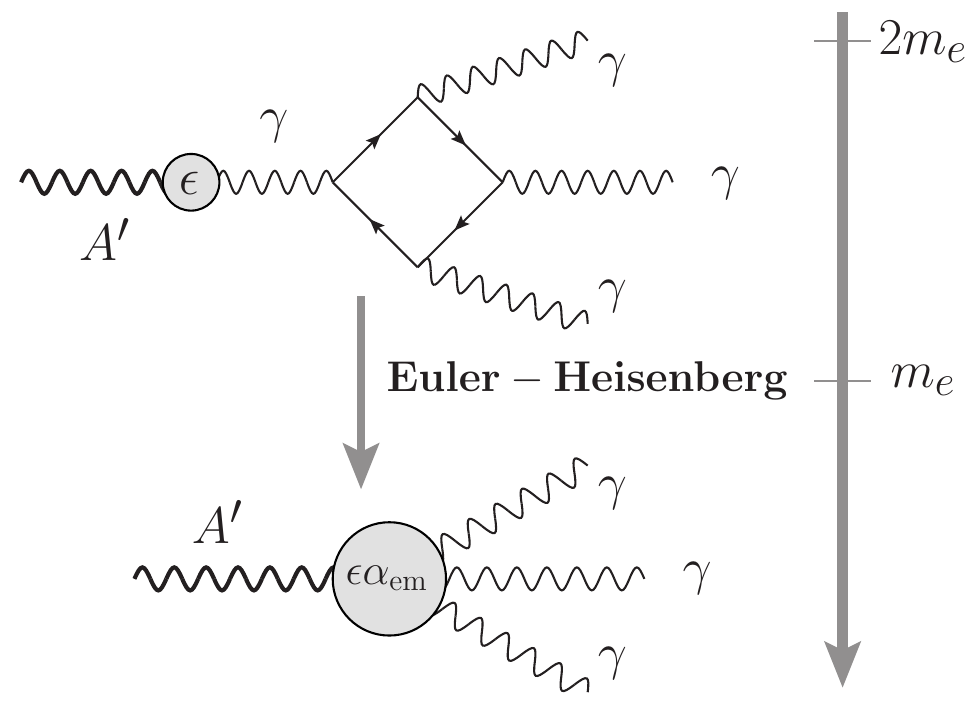}
\caption{Schematic illustration showing how the 
Euler-Heisenberg effective field theory (bottom) results from integrating out loops of SM charged fermions (top) in the limit $m_A \ll 2m_e$.}
\label{fig:diagram}
\end{figure}

However, INTEGRAL data analysis requires the removal of a isotropic emission sources, which are dominated by cosmic-ray contamination. Thus, we remove the isotropic portion of the dark matter annihilation signal setting the D-factor over the most distant region of our ROI (47.5$^\circ$) to 0 and renormalize the remaining dark matter signal. This leaves us with an effective D-factor of \mbox{$D=1.2\times 10^{23}$~GeV/cm$^{2}$}, which we use for our analysis. We note that this choice also effectively removes any isotropic x-ray background from extragalactic sources from our analysis~\cite{Iguaz:2021irx}.

We do not consider dark photon masses above $2 m_e$.  Despite the fact that such heavier dark photons can produce a significant photon flux from Final State Radiation (FSR)~\cite{Essig:2013goa}, this region is already stringently constrained by beam dump experiments~\cite{Essig:2013lka}, and model-dependent constraints from the CMB~\cite{Ellis:1990nb} and BBN~\cite{Iocco:2008va, Holtmann:1998gd, Kawasaki:2000qr}.

\vspace{5pt}


\noindent\textbf{\emph{INTEGRAL/SPI anaylysis.}} --- The INTErnational Gamma-Ray Astro Physics Laboratory, or INTEGRAL, is a hard x-ray space telescope. Using the SPectrometer of INTEGRAL, or SPI, it uniquely probes an important energy range that lies between 30~keV and 8~MeV. INTEGRAL observations open many opportunities in the field of astrophysics~\cite{Siegert:2015ila, Siegert:2015knp, Siegert:2016ymf, Siegert:2016ijv, Siegert:2019clp, Siegert:2021trw}, as well as searches for novel phenomena~\cite{Calore:2022pks, DelaTorreLuque:2024zsr, Berteaud:2022tws, Siegert:2024hmr, Iguaz:2021irx, Siegert:2021upf}.



Following the framework of previous studies focused on decays of axions and sterile neutrinos~\cite{Calore:2022pks}, as well as primordial black holes~\cite{Berteaud:2022tws}, we study the dark photon trident with 16 years of INTEGRAL/SPI data. Since we are interested in photons with energies below 511 keV, the diffuse background model we consider in this study contains three main components~\cite{Siegert:2019cvk}, which are shown in Fig.~\ref{fig:flux}:
\begin{itemize}
    \item Unresolved point sources (\textcolor{Red}{red bands}) described by a power-law with an exponential cutoff~\cite{unresolve};
    \item Inverse Compton scattering of interstellar radiation fields described by a power-law spectrum (\textcolor{Green}{green bands})~\cite{Wang_2020};
    \item Emission produced by positronium decays are in (\textcolor{cyan}{cyan bands}) with a normalization that is strongly constrained by 511 keV line observations~\cite{Siegert:2015knp}.
\end{itemize}

\noindent Additionally, there is a \textcolor{orange}{nuclear line} at 478 keV from $^{7}$Be, but its contribution is negligible~\cite{Siegert:2021wlq}. Analytic spectra for these components are shown in the Supplemental Material. The total diffuse flux is the combination of these components, shown at 1$\sigma$ and 2$\sigma$ in the \textcolor[HTML]{4B0082}{purple bands}. Figure~\ref{fig:flux} also depicts a 2-sigma uncertainty band for 400 keV dark photon decays as the \textcolor{gray}{gray band}, which has a peak around $m_{A^{\prime}}/3$, with the spectral shape from Eq.~\ref{eq:spectrum}. The normalization of this component is arbitrary and not included in the total diffuse fit.

We consider photon energies between 30--511~keV. We set a 95\% C.L. upper limit on the dark matter decay timescale by fitting the dark matter component along with the arbitrary normalizations of each astrophysical component, to INTEGRAL/SPI data. Using the {\bf 3ML} package~\cite{Vianello:2015wwa} with the \texttt{emcee} MCMC method~\cite{Foreman_Mackey_2013}, we sample all free spectral parameters, including the decay width of the dark photon trident. We utilize prior distributions that prevent un-physical results (e.g., negative fluxes from any source class). The template and codes used in our analysis are publicly available, and stem from previous work on dark matter decays in Refs.\cite{Calore:2022pks, Berteaud:2022tws}, published on the \href{https://zenodo.org/record/7984451}{Zendono} repository. \\

\begin{figure}[!t]
\centering
\includegraphics[width=\columnwidth]{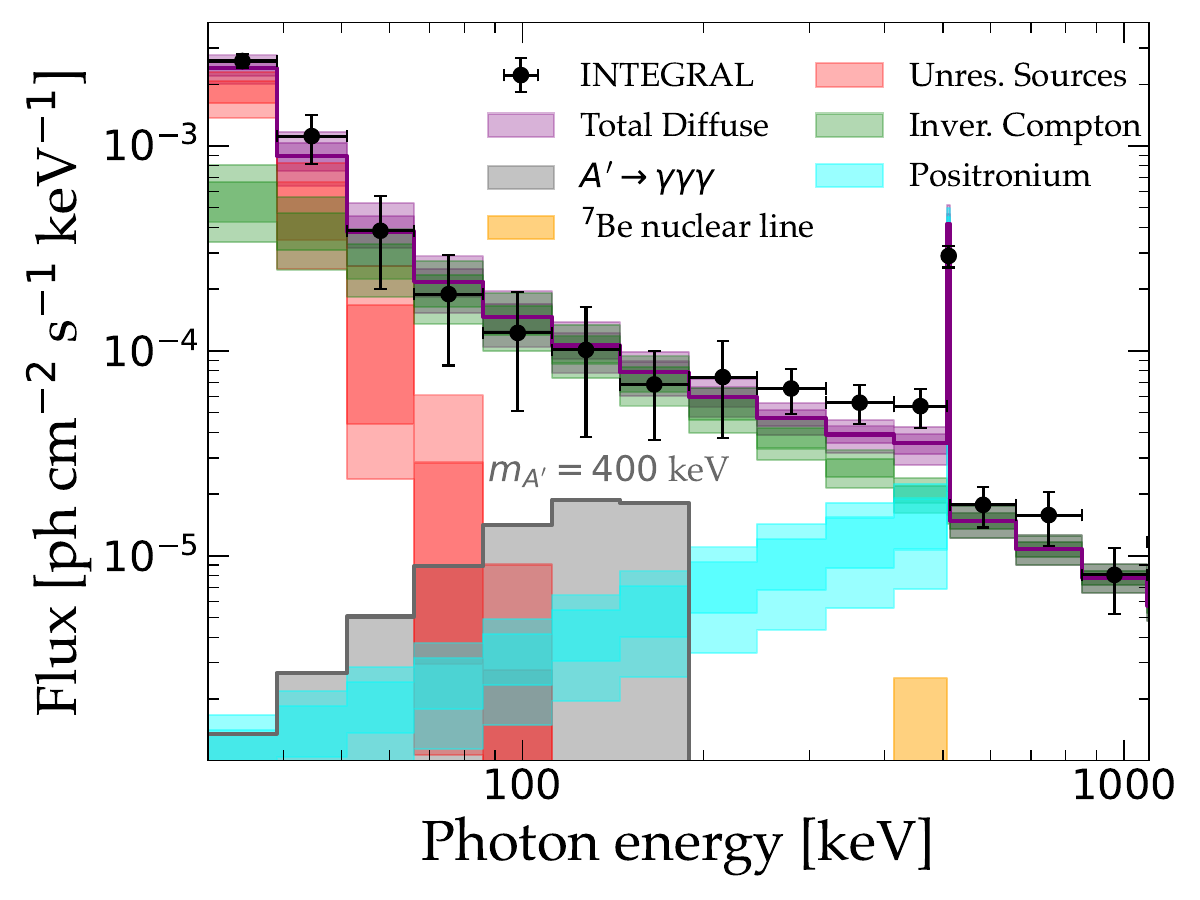}
\caption{The INTEGRAL photon flux and models for different sources. The observed data between 30--1022 keV are shown as black points. The best fit, along with 1$\sigma$ and 2$\sigma$ error bands for the astrophysical source classes are depicted as follows: \textcolor{Red}{Red} for \textcolor{Red}{Unresolved Sources}, \textcolor{Green}{green} for \textcolor{Green}{Inverse Compton} scattering, \textcolor{cyan}{Positronium} decay with an intensity fit to the 511~keV line is shown in \textcolor{cyan}{cyan}, and $^{7}{\rm Be}$ \textcolor{orange}{nuclear line} is in \textcolor{orange}{orange}. The \textcolor[HTML]{4B0082}{Total Diffuse Flux} that fits the data is in \textcolor[HTML]{4B0082}{purple}. For comparison, we also show an arbitrarily normalized spectrum for the  dark photon trident from a 400 keV dark photon (\textcolor{gray}{gray band}), which is not included in the fit.}
\label{fig:flux}
\end{figure}

\noindent\textbf{\emph{Constraints on dark photon dark matter.}} --- In Figure~\ref{fig:lifetime} (\textcolor{blue}{blue}), we show the 95\% C.L. upper limits on the dark matter lifetime in the case of dark photon trident decay using 16 years of INTEGRAL/SPI data. We compare our result with two existing constraints: dark photon direct detection results from Xenon in \textcolor{gray}{gray}~\cite{XENON:2020rca}, and previous searches for the same dark photon trident from Refs.~\cite{Yuksel:2007dr, Redondo:2008ec} in \textcolor{magenta}{magenta}. Our constraint also applies to other spin-1 dark matter models with a $\sim 100\%$ branching ratio into the three photons final states.



We map the constraints on the dark photon lifetime into constraints on the maximum allowed kinetic mixing parameter (using the one-loop decay amplitude).  These constraints are shown as the \textcolor{blue}{blue line} in Fig.~\ref{fig:constraints}. Direct Detection constraints from Xenon and projections for LZ~\cite{Bloch:2016sjj, XENON:2019gfn, XENON:2020rca, XENON:2021qze} are indicated by the solid and dashed \textcolor{gray}{gray lines}. The constraints from the dark photon trident using the intergalactic diffuse background limit (in the E-H approximation) from Refs.~\cite{Redondo:2008ec, Arias:2012az} are the solid \textcolor{magenta}{magenta line}.  The constraint that they would have obtained from the one-loop amplitude is shown as the dashed \textcolor{magenta}{magenta line}. 

We also include the combined constraints from numerous studies of dark photon production in stellar environments as the \textcolor{Green}{green line}, which can reach comparable strengths as LZ at low dark photon masses~\cite{Fischbach:1994ir, Zechlin:2008tj, Bi:2020ths, Li:2023vpv, Wadekar:2019mpc, Dolan:2023cjs, Dubovsky:2015cca, Yan:2023kdg, Hong:2020bxo, Vinyoles:2015aba}. We do note that, unlike the other constraints in this paper, stellar bound constraints have the advantage that they do not assume that the dark photon composes all (or even any) of the dark matter. However, their range is limited by the small plasma temperature of stellar objects, which limits the ability for stars to produce dark photons with masses above a few tens of keV~\cite{An:2013yfc}. Thus, our study demonstrates the unique power of diffuse x-ray data to probe a unique dark matter mass range where other techniques are insensitive.  




We finally compare our results with previous diffuse x-ray searches for dark photon trident decays from Refs.~\cite{Redondo:2008ec, Arias:2012az}. We note that our constraints significantly exceed previous work, even though both studies also used SPI~\cite{Churazov:2006bk} data (in addition to COMPTEL~\cite{comptel}, and EGRET~\cite{Strong:2004de} data) in their analysis. The strength of our constraints can be explained by three factors, which are listed in order of decreasing importance:
\begin{itemize}
    \item we take into account astrophysical background contributions, while Refs.~\cite{Redondo:2008ec, Arias:2012az} used the benchmark from Ref.~\cite{Yuksel:2007dr} which conservatively assumed that the observed photon signal is entirely produced by dark matter;
    \item we use the full one-loop calculation, instead of the leading EFT approximation used in Ref.~\cite{Redondo:2008ec};
    \item we use 16 years of data calibration, while Ref.~\cite{Yuksel:2007dr} used only 6 years of SPI observation.
\end{itemize}
The improved background modeling is the primary effect, and itself improves our constraint by around 1 order of magnitude, compared to Refs.~\cite{Redondo:2008ec, Arias:2012az}, even if the full one-loop decay amplitude were to be employed in both studies. This not only places new constraints on the dark photon dark matter model, but also again demonstrates the important of understanding astrophysical backgrounds for dark matter searches.

Note that in this study we only examine dark photon masses below the electron-positron pair threshold. For larger masses, the dominant decay channel is into electron-positron pairs that can subsequently radiate photons via final state radiation. The energy of such photons can populate the INTEGRAL search range, and thus it may be possible to set limits on dark photons with higher masses using this analysis. Since the motivation of this paper is to focus on the dark photon trident process, we leave investigation of this possibility for future work~\cite{Nguyen:2024kwy, Nguyen:2025eva, Nguyen:2025tkl}.

\begin{figure}[!t]
\centering
\includegraphics[width=\columnwidth]{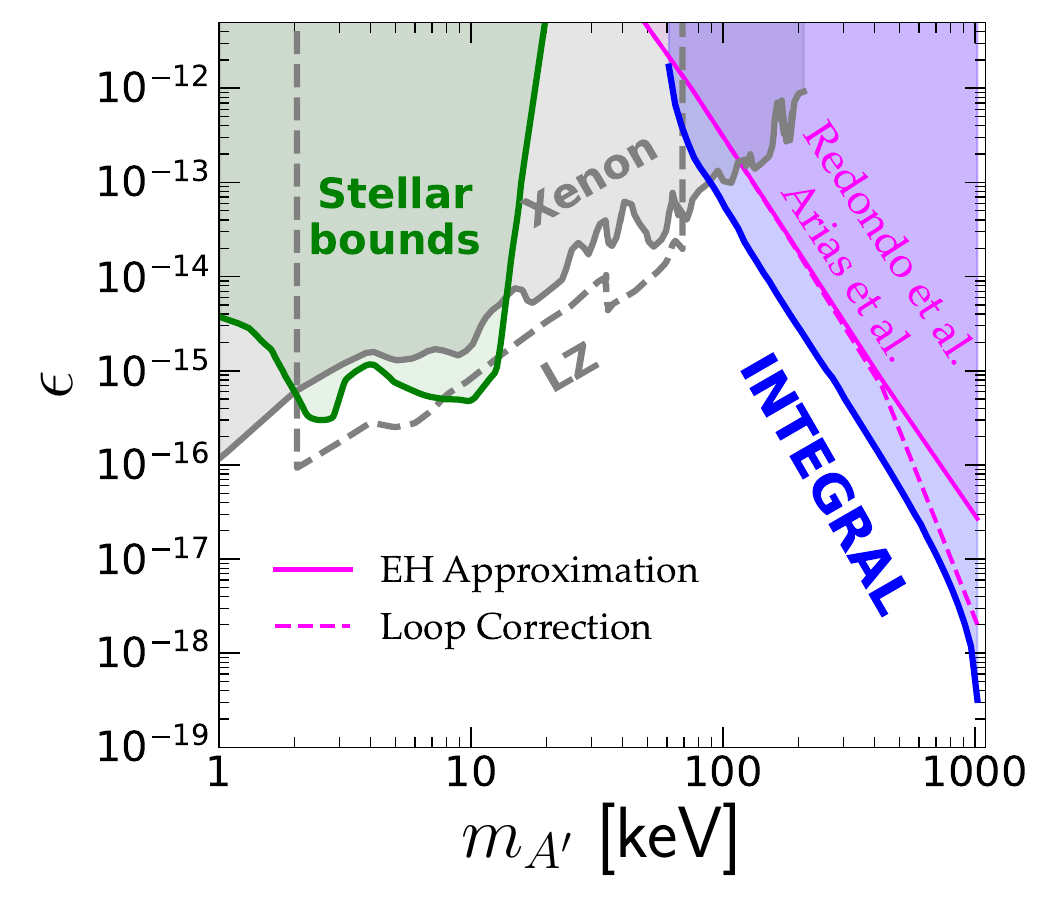}
\caption{New constraints on the plane of the kinetic mixing and the dark photon mass using \textcolor{blue}{INTEGRAL/SPI} are shown in \textcolor{blue}{blue}. \textcolor{magenta}{Previous constraints} based on this decay process based on the diffuse X-ray background is in \textcolor{magenta}{magenta}: solid line is the leading EFT result~\cite{Redondo:2008ec, Arias:2012az, Caputo:2021eaa}, and the dashed line includes the full one-loop decay amplitude. Constraints on dark photon from direct searches are shown in \textcolor{gray}{gray}: the solid line is from Xenon and the dashed line the projection for LZ~\cite{Bloch:2016sjj, XENON:2019gfn, XENON:2020rca, XENON:2021qze, An:2020bxd}. The combined constraints from studies of dark photon production in stars (\textcolor{Green}{stellar bounds}) are in \textcolor{Green}{green} and follow the results of Refs.~\cite{Fischbach:1994ir, Zechlin:2008tj, Bi:2020ths, Li:2023vpv, Wadekar:2019mpc, Dolan:2023cjs, Dubovsky:2015cca, Yan:2023kdg, Hong:2020bxo, Vinyoles:2015aba}.}
\label{fig:constraints}
\end{figure}

\vspace{5pt}

\noindent\textbf{\emph{Conclusion and outlook.}} --- In this {\it Letter}, we study the decay of dark photon dark matter using 16 years of INTEGRAL/SPI data in the mass range of 60--1022~keV. Concentrating on the case where the three photon final state (dark photon trident) is the dominant decay channel, we provide new constraints on kinetic mixing and dark photon mass that exceed previous results by nearly two orders of magnitude. Taking into account astrophysical background and the full one-loop decay amplitude, we constrain the kinetic mixing coupling from 10$^{-12}$ down to $10^{-19}$ in the 61-1022 keV mass range of dark photon. 

This study also motivates other searches for dark photon tridents. Since the dark photon is a weakly slim interacting particle (WISP) and has similar decay signal as the axion and axion-like-particles, many studies and techniques for WISP searches can also be applied to the dark photon~\cite{AxionLimits}. Our results motivate exploring existing x-ray and IR searches for axions, to see how they apply to the dark photon trident process, such as for example: Chandra~\cite{Wouters:2013hua, Marsh:2017yvc, Reynolds:2019uqt, Reynes:2021bpe}, XMM-Newton~\cite{Foster:2021ngm, DelaTorreLuque:2023nhh}, NuSTAR~\cite{Perez:2016tcq, Ng:2019gch, Roach:2022lgo}, eROSITA~\cite{Fong:2024qeq}, JWST~\cite{Janish:2023kvi, An:2024kls, Roy:2023omw, Bessho:2022yyu, Yin:2024lla}, MUSE~\cite{Todarello:2023hdk}, and future observations such as THESEUS~\cite{Thorpe-Morgan:2020rwc}, ASTROMEV~\cite{De_Angelis_2021}, XRISM~\cite{Dessert:2023vyl}, Athena~\cite{Barret:2013mxa, Sisk-Reynes:2022sqd},
Axis~\cite{Sisk-Reynes:2022sqd}, and EuXFEL~\cite{Halliday:2024lca}.

\vspace{3pt}

\noindent\textbf{\textit{Acknowledgements.}} --- We especially thank Francesca Calore and Thomas Siegert for sharing the modeling code from Ref.~\cite{Calore:2022pks}, which were used in this analysis. We also grateful for Ciaran O'Hare's comments on other existing constraints from \href{https://cajohare.github.io/AxionLimits/docs/dp.html}{his website}. We thank Carlos Blanco, Christopher Dessert, Ludvig Doeser, J{\'u}lia Sisk-Reyn{\'e}s, Nick Rodd, and Axel Widmark for fruitful discussions. T.L. and T.T.Q.N acknowledge support by the Swedish Research Council under contract 2022-04283. T.T.Q.N is also grateful for the supported by the COST Action COSMIC WISPers CA21106, supported by COST (European Cooperation in Science and Technology) and thank the Galileo Galilei Institute for Theoretical Physics for the hospitality and the INFN for partial support during the completion of this work. T.M.P.T. is supported by the US National Science Foundation under Grant PHY-2210283.

This work made use of {\tt Numpy}~\cite{Harris_2020}, {\tt SciPy}~\cite{Virtanen:2019joe}, {\tt astropy}~\cite{Astropy:2013muo}, {\tt matplotlib}~\cite{HunterMatplotlib}, {\tt Jupyter}~\cite{2016ppap.book...87K}, {\tt Jaxodraw}~\cite{Binosi:2008ig}, as well as {\tt Webplotdigitizer}~\cite{Rohatgi2022}.

\clearpage
\newpage
\maketitle
\onecolumngrid
\begin{center}
\textbf{\large X-Ray Constraints on Dark Photon Tridents}

\vspace{0.05in}
{ \it \large Supplemental Material}\\ 
\vspace{0.05in}
{Tim Linden, Thong T.Q. Nguyen, Tim M.P. Tait}
\end{center}
\onecolumngrid
\setcounter{equation}{0}
\setcounter{figure}{0}
\setcounter{section}{0}
\setcounter{table}{0}
\setcounter{page}{1}
\makeatletter
\renewcommand{\theequation}{S\arabic{equation}}
\renewcommand{\thefigure}{S\arabic{figure}}
\renewcommand{\thetable}{S\arabic{table}} 

\section{Dark photon trident decay width and energy spectrum}
We consider a dark photon decay into three photons through a loop induced by a fermion in Fig.~\ref{fig:diagram}, with the most significant contribution coming from the electron. The resulting amplitude can be parameterized as:
\begin{equation}
    i\mathcal{M}\equiv i\mathcal{M}(p, k_{1}, k_{2}, k_{3}, h, s_{1}, s_{2}, s_{3})=i\mathcal{A}^{\mu\nu\rho\sigma}\epsilon(p, h)\epsilon^{*}(k_{1}, s_{1})\epsilon^{*}(k_{2}, s_{2})\epsilon^{*}(k_{3}, s_{3}),
\end{equation}
where ($p$, $h$) and ($k_{i}$, $s_{i}$) are the momenta and polarizations of the initial dark photon and each final $i$th photon. For loop momentum $q$, the contribution from a single fermion to $\mathcal{A}^{\mu\nu\rho\sigma}$ is
\begin{equation}
    \begin{split}
        i\mathcal{A}^{\mu\nu\rho\sigma}=&2\epsilon e^{4}\Big{[}\int\frac{{\rm d}^{4}q}{(2\pi)^{4}}\frac{{\rm Tr}[\gamma^{\mu}(\slashed{q}+m_{e})\gamma^{\nu}(\slashed{q}+\slashed{k}_{1}+m_{e})\gamma^{\rho}(\slashed{q}+\slashed{k}_{1}+\slashed{k}_{2}+m_{e})\gamma^{\sigma}(\slashed{q}+\slashed{p}+m_{e})]}{(q^{2}+m_{e}^{2})[(q+k_{1})^{2}+m_{e}^{2}][(q+k_{1}+k_{2})^{2}+m_{e}^{2}][(q+p)^{2}-m_{e}^{2}]}\\
        &+\int\frac{{\rm d}^{4}q}{(2\pi)^{4}}\frac{{\rm Tr}[\gamma^{\mu}(\slashed{q}+m_{e})\gamma^{\rho}(\slashed{q}+\slashed{k}_{2}+m_{e})\gamma^{\nu}(\slashed{q}+\slashed{k}_{1}+\slashed{k}_{2}+m_{e})\gamma^{\sigma}(\slashed{q}+\slashed{p}+m_{e})]}{(q^{2}+m_{e}^{2})[(q+k_{2})^{2}+m_{e}^{2}][(q+k_{1}+k_{2})^{2}+m_{e}^{2}][(q+p)^{2}-m_{e}^{2}]}\\
        &+\int\frac{{\rm d}^{4}q}{(2\pi)^{4}}\frac{{\rm Tr}[\gamma^{\mu}(\slashed{q}+m_{e})\gamma^{\nu}(\slashed{q}+\slashed{k}_{1}+m_{e})\gamma^{\sigma}(\slashed{q}+\slashed{p}-\slashed{k}_{2}+m_{e})\gamma^{\rho}(\slashed{q}+\slashed{p}+m_{e})]}{(q^{2}+m_{e}^{2})[(q+k_{1})^{2}+m_{e}^{2}][(q+p-k_{2})^{2}+m_{e}^{2}][(q+p)^{2}-m_{e}^{2}]}\Big{]},
    \end{split}
    \label{eq:loop}
\end{equation}
where $m_e$ is the mass of the fermion.
The squared amplitude, summed over all final state polarizations and averaged over the initial 3 polarization of massive dark photon, is
\begin{equation}
\overline{|\mathcal{M}|}^{2}=\frac{1}{3}\mathcal{A}^{\mu\nu\rho\sigma}(\mathcal{A}^{*})^{\dot{\mu}\dot{\nu}\dot{\rho}\dot{\sigma}}\Big{(}-g_{\mu\dot{\mu}}+\frac{p_{\mu}p_{\dot{\mu}}}{m_{A^{\prime}}^{2}}\Big{)}(-g_{\nu\dot{\nu}})(-g_{\rho\dot{\rho}})(-g_{\sigma\dot{\sigma}})=\frac{1}{3}\mathcal{A}^{\mu\nu\rho\sigma}(\mathcal{A}^{*})_{\mu\nu\rho\sigma}.
\label{eq:M2}
\end{equation}

The dark photon trident decay width at one-loop is:
\begin{equation}
    \Gamma_{A^{\prime}\to 3\gamma}=\frac{1}{(2\pi)^{3}}\frac{1}{32m_{A^{\prime}}^{2}}\frac{1}{6}\int_{0}^{m_{A^{\prime}}^{2}}{\rm d}m_{12}^{2}\int_{0}^{m_{A^{\prime}}^{2}-m_{12}^{2}}{\rm d}m_{13}^{2}\times\overline{|\mathcal{M}|}^{2},
    \label{eq:Gamma_loop}
\end{equation}
with kinematic variables $m_{ij}^{2}=(k_{i}+k_{j})^{2}$, and the factor $1/6$ is for three identical particles in the final state. Comparing this result to the Euler-Heisenberg approximation in Eq.~\ref{eq:GEH}, this decay width can be parameterized following Ref.~\cite{McDermott:2017qcg}:
\begin{equation}
    \Gamma_{A^{\prime}\to 3\gamma}=\Gamma_{\rm EH}\times f_{\rm loop}(m_{A^{\prime}})=\Gamma_{\rm EH}\Big{[}1+\sum\limits_{n=1}^{\infty}c_{k}\Big{(}\frac{m_{A^{\prime}}^{2}}{m_{e}^{2}}\Big{)}^{n}\Big{]},
    \label{eq:ratio}
\end{equation}
where the coefficients $c_{n}$ which encode the ratio between the full one-loop and leading EFT approximation are calculated in Ref.~\cite{McDermott:2017qcg}.

The energy spectrum in Eq.~\ref{eq:flux} is proportional to the decay width differential in the energy of one photon~\cite{Pospelov:2007mp}:
\begin{equation}
    \frac{1}{N}\frac{{\rm d}N}{{\rm d}x}=\frac{1}{\Gamma_{A^{\prime}\to 3\gamma}}\frac{{\rm d}\Gamma_{A^{\prime}\to 3\gamma}}{{\rm d}x}=\frac{1}{\Gamma_{\rm EH}}\frac{{\rm d}\Gamma_{\rm EH}}{{\rm d}x}=\frac{1}{51}x^{3}\Big{(}1715-3105x+\frac{2919}{2}x^{2}\Big{)},
    \label{eq:dndx}
\end{equation}
where $x=2E_{\gamma}/m_{A^{\prime}}$. Notably, the energy spectrum is independent of the dark photon mass.

\section{Diffuse Astrophysical background}

To model the astrophysical background, we consider three main components that produce diffuse photons below 1~MeV:

\begin{itemize}
    \item A contribution from unresolved sources that is modeld as a power-law spectrum with an exponential cutoff given by:
    
    \begin{equation}
        \Phi_{\rm unres}=C_{0}\Big{(}\frac{E}{E_{0}}\Big{)}^{\alpha_{0}}\exp\Big{(}-\frac{E}{E_{C}}\Big{)}.
    \end{equation}
    \item A contribution from inverse-Compton scattering which has a power-law spectrum given by:
    \begin{equation}
        \Phi_{\rm IC}=C_{1}\Big{(}\frac{E}{E_{1}}\Big{)}^{\alpha_{1}}.
    \end{equation}
    \item A positronium decay spectrum with an amplitude that is fit to the flux of the 511~keV line, $\Phi_{\rm posi}=F(E, f_{\rm Ps}, F_{511})$, that is described in Ref.~\cite{Siegert:2016ymf}.
    \item Nuclear lines emission, the most important contribution for which stems from $^{7}$Be. This spectrum is fit by a Gaussian distribution
    \begin{equation}
        \Phi_{{\rm Be}}=\frac{F_{{\rm Be}}}{\sqrt{2\pi}\sigma_{\rm Be}}\exp\Big{(}-\frac{(E-\mu_{\rm Be})^{2}}{2\sigma_{\rm Be}^{2}}\Big{)}.
    \end{equation}

    \noindent where the Gaussian width corresponds primarily to the instrumental energy resolution.
\end{itemize}

These spectra are imported from the {\bf 3ML} package~\cite{Vianello:2015wwa}. All parameters' values and prior distributions are listed in Tab.~\ref{tab:parameter}.

\begin{table}[!h]
\centering
\renewcommand{\arraystretch}{1.5} 
\setlength{\tabcolsep}{8pt}
\begin{tabular}{|c|c|c|c|c|c|}
\hline
              Sources (Spectrum)    & Parameter & Fix/Free & Value/Range & Unit & Prior distribution \\ \hline
\multirow{4}{*}{Unresolved Sources} & $C_{0}$ & Free & $10^{-5}$--1.0 & keV$^{-1}$ s$^{-1}$ cm$^{-2}$ & Log Uniform\\ \cline{2-6} 
                  & $E_{0}$ & Fix & 50 & keV & \_ \\ \cline{2-6} 
                 & $\alpha_{0}$ & Fix & 0.0 & \_ & \_ \\ \cline{2-6} 
                  & $E_{C}$ & Free & 1--100 & keV & Truncated Gaussian \\ \hline
\multirow{3}{*}{Inverse Compton} & $C_{1}$ & Free & $10^{-10}$--1.0 & keV$^{-1}$ s$^{-1}$ cm$^{-2}$ & Log Uniform \\ \cline{2-6} 
                 & $E_{1}$ & Fix & 1000 & keV & \_ \\ \cline{2-6} 
                  & $\alpha_{1}$ & Free & -3.0--0.0 & \_ & Uniform \\ \hline
\multirow{2}{*}{Positronium} & $F_{511}$ & Free & $10^{-6}$--0.1 & s$^{-1}$ cm$^{-2}$ & Log Uniform \\ \cline{2-6} 
                  & $f_{\rm Ps}$ & Free & 0.0--1.0 & \_ & Truncated Gaussian \\ \hline
\multirow{3}{*}{Nuclear line ($^{7}$Be)} & $F_{\rm Be}$ & Free & $10^{-10}$--1.0 & s$^{-1}$ cm$^{-2}$ & Log Uniform \\ \cline{2-6} 
                  & $\sigma_{\rm Be}$ & Fix & 2.4 & keV & \_ \\ \cline{2-6} 
                  & $\mu_{\rm Be}$ & Fix & 478 & keV & \_ \\ \hline
\end{tabular}
\caption{Parameter values, scan ranges and their prior distributions in our \texttt{emcee} MCMC sampling priors for the astrophysical background, and are encoded in the {\bf 3ML} package. See also Ref.~\cite{Siegert:2024hmr}.}
\label{tab:parameter}
\end{table}

\bibliography{main}

\begin{thebibliography}{102}%
\makeatletter
\providecommand \@ifxundefined [1]{%
 \@ifx{#1\undefined}
}%
\providecommand \@ifnum [1]{%
 \ifnum #1\expandafter \@firstoftwo
 \else \expandafter \@secondoftwo
 \fi
}%
\providecommand \@ifx [1]{%
 \ifx #1\expandafter \@firstoftwo
 \else \expandafter \@secondoftwo
 \fi
}%
\providecommand \natexlab [1]{#1}%
\providecommand \enquote  [1]{``#1''}%
\providecommand \bibnamefont  [1]{#1}%
\providecommand \bibfnamefont [1]{#1}%
\providecommand \citenamefont [1]{#1}%
\providecommand \href@noop [0]{\@secondoftwo}%
\providecommand \href [0]{\begingroup \@sanitize@url \@href}%
\providecommand \@href[1]{\@@startlink{#1}\@@href}%
\providecommand \@@href[1]{\endgroup#1\@@endlink}%
\providecommand \@sanitize@url [0]{\catcode `\\12\catcode `\$12\catcode `\&12\catcode `\#12\catcode `\^12\catcode `\_12\catcode `\%12\relax}%
\providecommand \@@startlink[1]{}%
\providecommand \@@endlink[0]{}%
\providecommand \url  [0]{\begingroup\@sanitize@url \@url }%
\providecommand \@url [1]{\endgroup\@href {#1}{\urlprefix }}%
\providecommand \urlprefix  [0]{URL }%
\providecommand \Eprint [0]{\href }%
\providecommand \doibase [0]{http://dx.doi.org/}%
\providecommand \selectlanguage [0]{\@gobble}%
\providecommand \bibinfo  [0]{\@secondoftwo}%
\providecommand \bibfield  [0]{\@secondoftwo}%
\providecommand \translation [1]{[#1]}%
\providecommand \BibitemOpen [0]{}%
\providecommand \bibitemStop [0]{}%
\providecommand \bibitemNoStop [0]{.\EOS\space}%
\providecommand \EOS [0]{\spacefactor3000\relax}%
\providecommand \BibitemShut  [1]{\csname bibitem#1\endcsname}%
\let\auto@bib@innerbib\@empty
\bibitem [{\citenamefont {Holdom}(1986)}]{Holdom:1985ag}%
  \BibitemOpen
  \bibfield  {author} {\bibinfo {author} {\bibfnamefont {B.}~\bibnamefont {Holdom}},\ }\href {\doibase 10.1016/0370-2693(86)91377-8} {\bibfield  {journal} {\bibinfo  {journal} {Phys. Lett. B}\ }\textbf {\bibinfo {volume} {166}},\ \bibinfo {pages} {196} (\bibinfo {year} {1986})}\BibitemShut {NoStop}%
\bibitem [{\citenamefont {Fayet}(1980)}]{Fayet:1980ad}%
  \BibitemOpen
  \bibfield  {author} {\bibinfo {author} {\bibfnamefont {P.}~\bibnamefont {Fayet}},\ }\href {\doibase 10.1016/0370-2693(80)90488-8} {\bibfield  {journal} {\bibinfo  {journal} {Phys. Lett. B}\ }\textbf {\bibinfo {volume} {95}},\ \bibinfo {pages} {285} (\bibinfo {year} {1980})}\BibitemShut {NoStop}%
\bibitem [{\citenamefont {Fayet}(1990)}]{Fayet:1990wx}%
  \BibitemOpen
  \bibfield  {author} {\bibinfo {author} {\bibfnamefont {P.}~\bibnamefont {Fayet}},\ }\href {\doibase 10.1016/0550-3213(90)90381-M} {\bibfield  {journal} {\bibinfo  {journal} {Nucl. Phys. B}\ }\textbf {\bibinfo {volume} {347}},\ \bibinfo {pages} {743} (\bibinfo {year} {1990})}\BibitemShut {NoStop}%
\bibitem [{\citenamefont {Fabbrichesi}\ \emph {et~al.}(2020)\citenamefont {Fabbrichesi}, \citenamefont {Gabrielli},\ and\ \citenamefont {Lanfranchi}}]{Fabbrichesi:2020wbt}%
  \BibitemOpen
  \bibfield  {author} {\bibinfo {author} {\bibfnamefont {M.}~\bibnamefont {Fabbrichesi}}, \bibinfo {author} {\bibfnamefont {E.}~\bibnamefont {Gabrielli}}, \ and\ \bibinfo {author} {\bibfnamefont {G.}~\bibnamefont {Lanfranchi}},\ }\href {\doibase 10.1007/978-3-030-62519-1} {\  (\bibinfo {year} {2020}),\ 10.1007/978-3-030-62519-1},\ \Eprint {http://arxiv.org/abs/2005.01515} {arXiv:2005.01515 [hep-ph]} \BibitemShut {NoStop}%
\bibitem [{\citenamefont {Rizzo}(2018)}]{Rizzo:2018ntg}%
  \BibitemOpen
  \bibfield  {author} {\bibinfo {author} {\bibfnamefont {T.~G.}\ \bibnamefont {Rizzo}},\ }\href {\doibase 10.1007/JHEP07(2018)118} {\bibfield  {journal} {\bibinfo  {journal} {JHEP}\ }\textbf {\bibinfo {volume} {07}},\ \bibinfo {pages} {118} (\bibinfo {year} {2018})},\ \Eprint {http://arxiv.org/abs/1801.08525} {arXiv:1801.08525 [hep-ph]} \BibitemShut {NoStop}%
\bibitem [{\citenamefont {Pospelov}(2009)}]{Pospelov:2008zw}%
  \BibitemOpen
  \bibfield  {author} {\bibinfo {author} {\bibfnamefont {M.}~\bibnamefont {Pospelov}},\ }\href {\doibase 10.1103/PhysRevD.80.095002} {\bibfield  {journal} {\bibinfo  {journal} {Phys. Rev. D}\ }\textbf {\bibinfo {volume} {80}},\ \bibinfo {pages} {095002} (\bibinfo {year} {2009})},\ \Eprint {http://arxiv.org/abs/0811.1030} {arXiv:0811.1030 [hep-ph]} \BibitemShut {NoStop}%
\bibitem [{\citenamefont {Nelson}\ and\ \citenamefont {Scholtz}(2011)}]{Nelson:2011sf}%
  \BibitemOpen
  \bibfield  {author} {\bibinfo {author} {\bibfnamefont {A.~E.}\ \bibnamefont {Nelson}}\ and\ \bibinfo {author} {\bibfnamefont {J.}~\bibnamefont {Scholtz}},\ }\href {\doibase 10.1103/PhysRevD.84.103501} {\bibfield  {journal} {\bibinfo  {journal} {Phys. Rev. D}\ }\textbf {\bibinfo {volume} {84}},\ \bibinfo {pages} {103501} (\bibinfo {year} {2011})},\ \Eprint {http://arxiv.org/abs/1105.2812} {arXiv:1105.2812 [hep-ph]} \BibitemShut {NoStop}%
\bibitem [{\citenamefont {Caputo}\ \emph {et~al.}(2021)\citenamefont {Caputo}, \citenamefont {O'Hare}, \citenamefont {Millar},\ and\ \citenamefont {Vitagliano}}]{Caputo:2021eaa}%
  \BibitemOpen
  \bibfield  {author} {\bibinfo {author} {\bibfnamefont {A.}~\bibnamefont {Caputo}}, \bibinfo {author} {\bibfnamefont {C.~A.~J.}\ \bibnamefont {O'Hare}}, \bibinfo {author} {\bibfnamefont {A.~J.}\ \bibnamefont {Millar}}, \ and\ \bibinfo {author} {\bibfnamefont {E.}~\bibnamefont {Vitagliano}},\ }\href@noop {} {\  (\bibinfo {year} {2021})},\ \Eprint {http://arxiv.org/abs/2105.04565} {arXiv:2105.04565 [hep-ph]} \BibitemShut {NoStop}%
\bibitem [{\citenamefont {Redondo}\ and\ \citenamefont {Postma}(2009)}]{Redondo:2008ec}%
  \BibitemOpen
  \bibfield  {author} {\bibinfo {author} {\bibfnamefont {J.}~\bibnamefont {Redondo}}\ and\ \bibinfo {author} {\bibfnamefont {M.}~\bibnamefont {Postma}},\ }\href {\doibase 10.1088/1475-7516/2009/02/005} {\bibfield  {journal} {\bibinfo  {journal} {JCAP}\ }\textbf {\bibinfo {volume} {02}},\ \bibinfo {pages} {005} (\bibinfo {year} {2009})},\ \Eprint {http://arxiv.org/abs/0811.0326} {arXiv:0811.0326 [hep-ph]} \BibitemShut {NoStop}%
\bibitem [{\citenamefont {Arias}\ \emph {et~al.}(2012)\citenamefont {Arias}, \citenamefont {Cadamuro}, \citenamefont {Goodsell}, \citenamefont {Jaeckel}, \citenamefont {Redondo},\ and\ \citenamefont {Ringwald}}]{Arias:2012az}%
  \BibitemOpen
  \bibfield  {author} {\bibinfo {author} {\bibfnamefont {P.}~\bibnamefont {Arias}}, \bibinfo {author} {\bibfnamefont {D.}~\bibnamefont {Cadamuro}}, \bibinfo {author} {\bibfnamefont {M.}~\bibnamefont {Goodsell}}, \bibinfo {author} {\bibfnamefont {J.}~\bibnamefont {Jaeckel}}, \bibinfo {author} {\bibfnamefont {J.}~\bibnamefont {Redondo}}, \ and\ \bibinfo {author} {\bibfnamefont {A.}~\bibnamefont {Ringwald}},\ }\href {\doibase 10.1088/1475-7516/2012/06/013} {\bibfield  {journal} {\bibinfo  {journal} {JCAP}\ }\textbf {\bibinfo {volume} {06}},\ \bibinfo {pages} {013} (\bibinfo {year} {2012})},\ \Eprint {http://arxiv.org/abs/1201.5902} {arXiv:1201.5902 [hep-ph]} \BibitemShut {NoStop}%
\bibitem [{\citenamefont {Bloch}\ \emph {et~al.}(2017)\citenamefont {Bloch}, \citenamefont {Essig}, \citenamefont {Tobioka}, \citenamefont {Volansky},\ and\ \citenamefont {Yu}}]{Bloch:2016sjj}%
  \BibitemOpen
  \bibfield  {author} {\bibinfo {author} {\bibfnamefont {I.~M.}\ \bibnamefont {Bloch}}, \bibinfo {author} {\bibfnamefont {R.}~\bibnamefont {Essig}}, \bibinfo {author} {\bibfnamefont {K.}~\bibnamefont {Tobioka}}, \bibinfo {author} {\bibfnamefont {T.}~\bibnamefont {Volansky}}, \ and\ \bibinfo {author} {\bibfnamefont {T.-T.}\ \bibnamefont {Yu}},\ }\href {\doibase 10.1007/JHEP06(2017)087} {\bibfield  {journal} {\bibinfo  {journal} {JHEP}\ }\textbf {\bibinfo {volume} {06}},\ \bibinfo {pages} {087} (\bibinfo {year} {2017})},\ \Eprint {http://arxiv.org/abs/1608.02123} {arXiv:1608.02123 [hep-ph]} \BibitemShut {NoStop}%
\bibitem [{\citenamefont {Aprile}\ \emph {et~al.}(2019)\citenamefont {Aprile} \emph {et~al.}}]{XENON:2019gfn}%
  \BibitemOpen
  \bibfield  {author} {\bibinfo {author} {\bibfnamefont {E.}~\bibnamefont {Aprile}} \emph {et~al.} (\bibinfo {collaboration} {XENON}),\ }\href {\doibase 10.1103/PhysRevLett.123.251801} {\bibfield  {journal} {\bibinfo  {journal} {Phys. Rev. Lett.}\ }\textbf {\bibinfo {volume} {123}},\ \bibinfo {pages} {251801} (\bibinfo {year} {2019})},\ \Eprint {http://arxiv.org/abs/1907.11485} {arXiv:1907.11485 [hep-ex]} \BibitemShut {NoStop}%
\bibitem [{\citenamefont {Aprile}\ \emph {et~al.}(2020)\citenamefont {Aprile} \emph {et~al.}}]{XENON:2020rca}%
  \BibitemOpen
  \bibfield  {author} {\bibinfo {author} {\bibfnamefont {E.}~\bibnamefont {Aprile}} \emph {et~al.} (\bibinfo {collaboration} {XENON}),\ }\href {\doibase 10.1103/PhysRevD.102.072004} {\bibfield  {journal} {\bibinfo  {journal} {Phys. Rev. D}\ }\textbf {\bibinfo {volume} {102}},\ \bibinfo {pages} {072004} (\bibinfo {year} {2020})},\ \Eprint {http://arxiv.org/abs/2006.09721} {arXiv:2006.09721 [hep-ex]} \BibitemShut {NoStop}%
\bibitem [{\citenamefont {Aprile}\ \emph {et~al.}(2022)\citenamefont {Aprile} \emph {et~al.}}]{XENON:2021qze}%
  \BibitemOpen
  \bibfield  {author} {\bibinfo {author} {\bibfnamefont {E.}~\bibnamefont {Aprile}} \emph {et~al.} (\bibinfo {collaboration} {XENON}),\ }\href {\doibase 10.1103/PhysRevD.106.022001} {\bibfield  {journal} {\bibinfo  {journal} {Phys. Rev. D}\ }\textbf {\bibinfo {volume} {106}},\ \bibinfo {pages} {022001} (\bibinfo {year} {2022})},\ \Eprint {http://arxiv.org/abs/2112.12116} {arXiv:2112.12116 [hep-ex]} \BibitemShut {NoStop}%
\bibitem [{\citenamefont {An}\ \emph {et~al.}(2020)\citenamefont {An}, \citenamefont {Pospelov}, \citenamefont {Pradler},\ and\ \citenamefont {Ritz}}]{An:2020bxd}%
  \BibitemOpen
  \bibfield  {author} {\bibinfo {author} {\bibfnamefont {H.}~\bibnamefont {An}}, \bibinfo {author} {\bibfnamefont {M.}~\bibnamefont {Pospelov}}, \bibinfo {author} {\bibfnamefont {J.}~\bibnamefont {Pradler}}, \ and\ \bibinfo {author} {\bibfnamefont {A.}~\bibnamefont {Ritz}},\ }\href {\doibase 10.1103/PhysRevD.102.115022} {\bibfield  {journal} {\bibinfo  {journal} {Phys. Rev. D}\ }\textbf {\bibinfo {volume} {102}},\ \bibinfo {pages} {115022} (\bibinfo {year} {2020})},\ \Eprint {http://arxiv.org/abs/2006.13929} {arXiv:2006.13929 [hep-ph]} \BibitemShut {NoStop}%
\bibitem [{\citenamefont {Fischbach}\ \emph {et~al.}(1994)\citenamefont {Fischbach}, \citenamefont {Kloor}, \citenamefont {Langel}, \citenamefont {Liu},\ and\ \citenamefont {Peredo}}]{Fischbach:1994ir}%
  \BibitemOpen
  \bibfield  {author} {\bibinfo {author} {\bibfnamefont {E.}~\bibnamefont {Fischbach}}, \bibinfo {author} {\bibfnamefont {H.}~\bibnamefont {Kloor}}, \bibinfo {author} {\bibfnamefont {R.~A.}\ \bibnamefont {Langel}}, \bibinfo {author} {\bibfnamefont {A.~T.~Y.}\ \bibnamefont {Liu}}, \ and\ \bibinfo {author} {\bibfnamefont {M.}~\bibnamefont {Peredo}},\ }\href {\doibase 10.1103/PhysRevLett.73.514} {\bibfield  {journal} {\bibinfo  {journal} {Phys. Rev. Lett.}\ }\textbf {\bibinfo {volume} {73}},\ \bibinfo {pages} {514} (\bibinfo {year} {1994})}\BibitemShut {NoStop}%
\bibitem [{\citenamefont {Zechlin}\ \emph {et~al.}(2009)\citenamefont {Zechlin}, \citenamefont {Horns},\ and\ \citenamefont {Redondo}}]{Zechlin:2008tj}%
  \BibitemOpen
  \bibfield  {author} {\bibinfo {author} {\bibfnamefont {H.-S.}\ \bibnamefont {Zechlin}}, \bibinfo {author} {\bibfnamefont {D.}~\bibnamefont {Horns}}, \ and\ \bibinfo {author} {\bibfnamefont {J.}~\bibnamefont {Redondo}},\ }\href {\doibase 10.1063/1.3076781} {\bibfield  {journal} {\bibinfo  {journal} {AIP Conf. Proc.}\ }\textbf {\bibinfo {volume} {1085}},\ \bibinfo {pages} {727} (\bibinfo {year} {2009})},\ \Eprint {http://arxiv.org/abs/0810.5501} {arXiv:0810.5501 [astro-ph]} \BibitemShut {NoStop}%
\bibitem [{\citenamefont {Bi}\ \emph {et~al.}(2021)\citenamefont {Bi}, \citenamefont {Gao}, \citenamefont {Guo}, \citenamefont {Houston}, \citenamefont {Li}, \citenamefont {Xu},\ and\ \citenamefont {Zhang}}]{Bi:2020ths}%
  \BibitemOpen
  \bibfield  {author} {\bibinfo {author} {\bibfnamefont {X.-J.}\ \bibnamefont {Bi}}, \bibinfo {author} {\bibfnamefont {Y.}~\bibnamefont {Gao}}, \bibinfo {author} {\bibfnamefont {J.}~\bibnamefont {Guo}}, \bibinfo {author} {\bibfnamefont {N.}~\bibnamefont {Houston}}, \bibinfo {author} {\bibfnamefont {T.}~\bibnamefont {Li}}, \bibinfo {author} {\bibfnamefont {F.}~\bibnamefont {Xu}}, \ and\ \bibinfo {author} {\bibfnamefont {X.}~\bibnamefont {Zhang}},\ }\href {\doibase 10.1103/PhysRevD.103.043018} {\bibfield  {journal} {\bibinfo  {journal} {Phys. Rev. D}\ }\textbf {\bibinfo {volume} {103}},\ \bibinfo {pages} {043018} (\bibinfo {year} {2021})},\ \Eprint {http://arxiv.org/abs/2002.01796} {arXiv:2002.01796 [astro-ph.HE]} \BibitemShut {NoStop}%
\bibitem [{\citenamefont {Tran}\ \emph {et~al.}(2024)\citenamefont {Tran}, \citenamefont {Nguyen},\ and\ \citenamefont {Yuan}}]{Tran:2023lzv}%
  \BibitemOpen
  \bibfield  {author} {\bibinfo {author} {\bibfnamefont {V.~Q.}\ \bibnamefont {Tran}}, \bibinfo {author} {\bibfnamefont {T.~T.~Q.}\ \bibnamefont {Nguyen}}, \ and\ \bibinfo {author} {\bibfnamefont {T.-C.}\ \bibnamefont {Yuan}},\ }\href {\doibase 10.1088/1475-7516/2024/05/015} {\bibfield  {journal} {\bibinfo  {journal} {JCAP}\ }\textbf {\bibinfo {volume} {05}},\ \bibinfo {pages} {015} (\bibinfo {year} {2024})},\ \Eprint {http://arxiv.org/abs/2312.10785} {arXiv:2312.10785 [hep-ph]} \BibitemShut {NoStop}%
\bibitem [{\citenamefont {Li}\ and\ \citenamefont {Xu}(2023)}]{Li:2023vpv}%
  \BibitemOpen
  \bibfield  {author} {\bibinfo {author} {\bibfnamefont {S.-P.}\ \bibnamefont {Li}}\ and\ \bibinfo {author} {\bibfnamefont {X.-J.}\ \bibnamefont {Xu}},\ }\href {\doibase 10.1088/1475-7516/2023/09/009} {\bibfield  {journal} {\bibinfo  {journal} {JCAP}\ }\textbf {\bibinfo {volume} {09}},\ \bibinfo {pages} {009} (\bibinfo {year} {2023})},\ \Eprint {http://arxiv.org/abs/2304.12907} {arXiv:2304.12907 [hep-ph]} \BibitemShut {NoStop}%
\bibitem [{\citenamefont {Wadekar}\ and\ \citenamefont {Farrar}(2021)}]{Wadekar:2019mpc}%
  \BibitemOpen
  \bibfield  {author} {\bibinfo {author} {\bibfnamefont {D.}~\bibnamefont {Wadekar}}\ and\ \bibinfo {author} {\bibfnamefont {G.~R.}\ \bibnamefont {Farrar}},\ }\href {\doibase 10.1103/PhysRevD.103.123028} {\bibfield  {journal} {\bibinfo  {journal} {Phys. Rev. D}\ }\textbf {\bibinfo {volume} {103}},\ \bibinfo {pages} {123028} (\bibinfo {year} {2021})},\ \Eprint {http://arxiv.org/abs/1903.12190} {arXiv:1903.12190 [hep-ph]} \BibitemShut {NoStop}%
\bibitem [{\citenamefont {Dolan}\ \emph {et~al.}(2023)\citenamefont {Dolan}, \citenamefont {Hiskens},\ and\ \citenamefont {Volkas}}]{Dolan:2023cjs}%
  \BibitemOpen
  \bibfield  {author} {\bibinfo {author} {\bibfnamefont {M.~J.}\ \bibnamefont {Dolan}}, \bibinfo {author} {\bibfnamefont {F.~J.}\ \bibnamefont {Hiskens}}, \ and\ \bibinfo {author} {\bibfnamefont {R.~R.}\ \bibnamefont {Volkas}},\ }\href@noop {} {\  (\bibinfo {year} {2023})},\ \Eprint {http://arxiv.org/abs/2306.13335} {arXiv:2306.13335 [hep-ph]} \BibitemShut {NoStop}%
\bibitem [{\citenamefont {Dubovsky}\ and\ \citenamefont {Hern\'andez-Chifflet}(2015)}]{Dubovsky:2015cca}%
  \BibitemOpen
  \bibfield  {author} {\bibinfo {author} {\bibfnamefont {S.}~\bibnamefont {Dubovsky}}\ and\ \bibinfo {author} {\bibfnamefont {G.}~\bibnamefont {Hern\'andez-Chifflet}},\ }\href {\doibase 10.1088/1475-7516/2015/12/054} {\bibfield  {journal} {\bibinfo  {journal} {JCAP}\ }\textbf {\bibinfo {volume} {12}},\ \bibinfo {pages} {054} (\bibinfo {year} {2015})},\ \Eprint {http://arxiv.org/abs/1509.00039} {arXiv:1509.00039 [hep-ph]} \BibitemShut {NoStop}%
\bibitem [{\citenamefont {Yan}\ \emph {et~al.}(2023)\citenamefont {Yan}, \citenamefont {Li},\ and\ \citenamefont {Fan}}]{Yan:2023kdg}%
  \BibitemOpen
  \bibfield  {author} {\bibinfo {author} {\bibfnamefont {S.}~\bibnamefont {Yan}}, \bibinfo {author} {\bibfnamefont {L.}~\bibnamefont {Li}}, \ and\ \bibinfo {author} {\bibfnamefont {J.}~\bibnamefont {Fan}},\ }\href@noop {} {\  (\bibinfo {year} {2023})},\ \Eprint {http://arxiv.org/abs/2312.06746} {arXiv:2312.06746 [hep-ph]} \BibitemShut {NoStop}%
\bibitem [{\citenamefont {Hong}\ \emph {et~al.}(2021)\citenamefont {Hong}, \citenamefont {Shin},\ and\ \citenamefont {Yun}}]{Hong:2020bxo}%
  \BibitemOpen
  \bibfield  {author} {\bibinfo {author} {\bibfnamefont {D.~K.}\ \bibnamefont {Hong}}, \bibinfo {author} {\bibfnamefont {C.~S.}\ \bibnamefont {Shin}}, \ and\ \bibinfo {author} {\bibfnamefont {S.}~\bibnamefont {Yun}},\ }\href {\doibase 10.1103/PhysRevD.103.123031} {\bibfield  {journal} {\bibinfo  {journal} {Phys. Rev. D}\ }\textbf {\bibinfo {volume} {103}},\ \bibinfo {pages} {123031} (\bibinfo {year} {2021})},\ \Eprint {http://arxiv.org/abs/2012.05427} {arXiv:2012.05427 [hep-ph]} \BibitemShut {NoStop}%
\bibitem [{\citenamefont {Vinyoles}\ \emph {et~al.}(2015)\citenamefont {Vinyoles}, \citenamefont {Serenelli}, \citenamefont {Villante}, \citenamefont {Basu}, \citenamefont {Redondo},\ and\ \citenamefont {Isern}}]{Vinyoles:2015aba}%
  \BibitemOpen
  \bibfield  {author} {\bibinfo {author} {\bibfnamefont {N.}~\bibnamefont {Vinyoles}}, \bibinfo {author} {\bibfnamefont {A.}~\bibnamefont {Serenelli}}, \bibinfo {author} {\bibfnamefont {F.~L.}\ \bibnamefont {Villante}}, \bibinfo {author} {\bibfnamefont {S.}~\bibnamefont {Basu}}, \bibinfo {author} {\bibfnamefont {J.}~\bibnamefont {Redondo}}, \ and\ \bibinfo {author} {\bibfnamefont {J.}~\bibnamefont {Isern}},\ }\href {\doibase 10.1088/1475-7516/2015/10/015} {\bibfield  {journal} {\bibinfo  {journal} {JCAP}\ }\textbf {\bibinfo {volume} {10}},\ \bibinfo {pages} {015} (\bibinfo {year} {2015})},\ \Eprint {http://arxiv.org/abs/1501.01639} {arXiv:1501.01639 [astro-ph.SR]} \BibitemShut {NoStop}%
\bibitem [{\citenamefont {Bertone}\ \emph {et~al.}(2005)\citenamefont {Bertone}, \citenamefont {Hooper},\ and\ \citenamefont {Silk}}]{Bertone:2004pz}%
  \BibitemOpen
  \bibfield  {author} {\bibinfo {author} {\bibfnamefont {G.}~\bibnamefont {Bertone}}, \bibinfo {author} {\bibfnamefont {D.}~\bibnamefont {Hooper}}, \ and\ \bibinfo {author} {\bibfnamefont {J.}~\bibnamefont {Silk}},\ }\href {\doibase 10.1016/j.physrep.2004.08.031} {\bibfield  {journal} {\bibinfo  {journal} {Phys. Rept.}\ }\textbf {\bibinfo {volume} {405}},\ \bibinfo {pages} {279} (\bibinfo {year} {2005})},\ \Eprint {http://arxiv.org/abs/hep-ph/0404175} {arXiv:hep-ph/0404175} \BibitemShut {NoStop}%
\bibitem [{\citenamefont {Bertone}\ and\ \citenamefont {Hooper}(2018)}]{Bertone:2016nfn}%
  \BibitemOpen
  \bibfield  {author} {\bibinfo {author} {\bibfnamefont {G.}~\bibnamefont {Bertone}}\ and\ \bibinfo {author} {\bibfnamefont {D.}~\bibnamefont {Hooper}},\ }\href {\doibase 10.1103/RevModPhys.90.045002} {\bibfield  {journal} {\bibinfo  {journal} {Rev. Mod. Phys.}\ }\textbf {\bibinfo {volume} {90}},\ \bibinfo {pages} {045002} (\bibinfo {year} {2018})},\ \Eprint {http://arxiv.org/abs/1605.04909} {arXiv:1605.04909 [astro-ph.CO]} \BibitemShut {NoStop}%
\bibitem [{\citenamefont {Bertone}\ and\ \citenamefont {Tait}(2018)}]{Bertone:2018krk}%
  \BibitemOpen
  \bibfield  {author} {\bibinfo {author} {\bibfnamefont {G.}~\bibnamefont {Bertone}}\ and\ \bibinfo {author} {\bibfnamefont {T.}~\bibnamefont {Tait}, \bibfnamefont {M.~P.}},\ }\href {\doibase 10.1038/s41586-018-0542-z} {\bibfield  {journal} {\bibinfo  {journal} {Nature}\ }\textbf {\bibinfo {volume} {562}},\ \bibinfo {pages} {51} (\bibinfo {year} {2018})},\ \Eprint {http://arxiv.org/abs/1810.01668} {arXiv:1810.01668 [astro-ph.CO]} \BibitemShut {NoStop}%
\bibitem [{\citenamefont {Servant}\ and\ \citenamefont {Tait}(2003)}]{Servant:2002aq}%
  \BibitemOpen
  \bibfield  {author} {\bibinfo {author} {\bibfnamefont {G.}~\bibnamefont {Servant}}\ and\ \bibinfo {author} {\bibfnamefont {T.~M.~P.}\ \bibnamefont {Tait}},\ }\href {\doibase 10.1016/S0550-3213(02)01012-X} {\bibfield  {journal} {\bibinfo  {journal} {Nucl. Phys. B}\ }\textbf {\bibinfo {volume} {650}},\ \bibinfo {pages} {391} (\bibinfo {year} {2003})},\ \Eprint {http://arxiv.org/abs/hep-ph/0206071} {arXiv:hep-ph/0206071} \BibitemShut {NoStop}%
\bibitem [{\citenamefont {Landau}(1948)}]{Landau:1948kw}%
  \BibitemOpen
  \bibfield  {author} {\bibinfo {author} {\bibfnamefont {L.~D.}\ \bibnamefont {Landau}},\ }\href {\doibase 10.1016/B978-0-08-010586-4.50070-5} {\bibfield  {journal} {\bibinfo  {journal} {Dokl. Akad. Nauk SSSR}\ }\textbf {\bibinfo {volume} {60}},\ \bibinfo {pages} {207} (\bibinfo {year} {1948})}\BibitemShut {NoStop}%
\bibitem [{\citenamefont {Yang}(1950)}]{Yang:1950rg}%
  \BibitemOpen
  \bibfield  {author} {\bibinfo {author} {\bibfnamefont {C.-N.}\ \bibnamefont {Yang}},\ }\href {\doibase 10.1103/PhysRev.77.242} {\bibfield  {journal} {\bibinfo  {journal} {Phys. Rev.}\ }\textbf {\bibinfo {volume} {77}},\ \bibinfo {pages} {242} (\bibinfo {year} {1950})}\BibitemShut {NoStop}%
\bibitem [{\citenamefont {Pospelov}\ \emph {et~al.}(2008)\citenamefont {Pospelov}, \citenamefont {Ritz},\ and\ \citenamefont {Voloshin}}]{Pospelov:2007mp}%
  \BibitemOpen
  \bibfield  {author} {\bibinfo {author} {\bibfnamefont {M.}~\bibnamefont {Pospelov}}, \bibinfo {author} {\bibfnamefont {A.}~\bibnamefont {Ritz}}, \ and\ \bibinfo {author} {\bibfnamefont {M.~B.}\ \bibnamefont {Voloshin}},\ }\href {\doibase 10.1016/j.physletb.2008.02.052} {\bibfield  {journal} {\bibinfo  {journal} {Phys. Lett. B}\ }\textbf {\bibinfo {volume} {662}},\ \bibinfo {pages} {53} (\bibinfo {year} {2008})},\ \Eprint {http://arxiv.org/abs/0711.4866} {arXiv:0711.4866 [hep-ph]} \BibitemShut {NoStop}%
\bibitem [{\citenamefont {Yuksel}\ and\ \citenamefont {Kistler}(2008)}]{Yuksel:2007dr}%
  \BibitemOpen
  \bibfield  {author} {\bibinfo {author} {\bibfnamefont {H.}~\bibnamefont {Yuksel}}\ and\ \bibinfo {author} {\bibfnamefont {M.~D.}\ \bibnamefont {Kistler}},\ }\href {\doibase 10.1103/PhysRevD.78.023502} {\bibfield  {journal} {\bibinfo  {journal} {Phys. Rev. D}\ }\textbf {\bibinfo {volume} {78}},\ \bibinfo {pages} {023502} (\bibinfo {year} {2008})},\ \Eprint {http://arxiv.org/abs/0711.2906} {arXiv:0711.2906 [astro-ph]} \BibitemShut {NoStop}%
\bibitem [{\citenamefont {Nguyen}\ and\ \citenamefont {Tait}(2023)}]{Nguyen:2022zwb}%
  \BibitemOpen
  \bibfield  {author} {\bibinfo {author} {\bibfnamefont {T.~T.~Q.}\ \bibnamefont {Nguyen}}\ and\ \bibinfo {author} {\bibfnamefont {T.~M.~P.}\ \bibnamefont {Tait}},\ }\href {\doibase 10.1103/PhysRevD.107.115016} {\bibfield  {journal} {\bibinfo  {journal} {Phys. Rev. D}\ }\textbf {\bibinfo {volume} {107}},\ \bibinfo {pages} {115016} (\bibinfo {year} {2023})},\ \Eprint {http://arxiv.org/abs/2212.12547} {arXiv:2212.12547 [hep-ph]} \BibitemShut {NoStop}%
\bibitem [{\citenamefont {Nguyen}\ \emph {et~al.}(2025{\natexlab{a}})\citenamefont {Nguyen}, \citenamefont {Linden}, \citenamefont {Carenza},\ and\ \citenamefont {Widmark}}]{Nguyen:2025ygc}%
  \BibitemOpen
  \bibfield  {author} {\bibinfo {author} {\bibfnamefont {T.~T.~Q.}\ \bibnamefont {Nguyen}}, \bibinfo {author} {\bibfnamefont {T.}~\bibnamefont {Linden}}, \bibinfo {author} {\bibfnamefont {P.}~\bibnamefont {Carenza}}, \ and\ \bibinfo {author} {\bibfnamefont {A.}~\bibnamefont {Widmark}},\ }\href@noop {} {\  (\bibinfo {year} {2025}{\natexlab{a}})},\ \Eprint {http://arxiv.org/abs/2501.14864} {arXiv:2501.14864 [astro-ph.HE]} \BibitemShut {NoStop}%
\bibitem [{\citenamefont {Nguyen}(2023)}]{Nguyen:2023ugx}%
  \BibitemOpen
  \bibfield  {author} {\bibinfo {author} {\bibfnamefont {T.~T.~Q.}\ \bibnamefont {Nguyen}},\ }in\ \href@noop {} {\emph {\bibinfo {booktitle} {{Windows on the Universe}: {30th Anniversary of the Rencontres du Vietnam}}}}\ (\bibinfo {year} {2023})\ \Eprint {http://arxiv.org/abs/2312.12292} {arXiv:2312.12292 [hep-ph]} \BibitemShut {NoStop}%
\bibitem [{\citenamefont {McDermott}\ \emph {et~al.}(2018)\citenamefont {McDermott}, \citenamefont {Patel},\ and\ \citenamefont {Ramani}}]{McDermott:2017qcg}%
  \BibitemOpen
  \bibfield  {author} {\bibinfo {author} {\bibfnamefont {S.~D.}\ \bibnamefont {McDermott}}, \bibinfo {author} {\bibfnamefont {H.~H.}\ \bibnamefont {Patel}}, \ and\ \bibinfo {author} {\bibfnamefont {H.}~\bibnamefont {Ramani}},\ }\href {\doibase 10.1103/PhysRevD.97.073005} {\bibfield  {journal} {\bibinfo  {journal} {Phys. Rev. D}\ }\textbf {\bibinfo {volume} {97}},\ \bibinfo {pages} {073005} (\bibinfo {year} {2018})},\ \Eprint {http://arxiv.org/abs/1705.00619} {arXiv:1705.00619 [hep-ph]} \BibitemShut {NoStop}%
\bibitem [{\citenamefont {Linden}\ \emph {et~al.}(2024)\citenamefont {Linden}, \citenamefont {Nguyen},\ and\ \citenamefont {Tait}}]{Linden:2024uph}%
  \BibitemOpen
  \bibfield  {author} {\bibinfo {author} {\bibfnamefont {T.}~\bibnamefont {Linden}}, \bibinfo {author} {\bibfnamefont {T.~T.~Q.}\ \bibnamefont {Nguyen}}, \ and\ \bibinfo {author} {\bibfnamefont {T.~M.~P.}\ \bibnamefont {Tait}},\ }\href@noop {} {\  (\bibinfo {year} {2024})},\ \Eprint {http://arxiv.org/abs/2402.01839} {arXiv:2402.01839 [hep-ph]} \BibitemShut {NoStop}%
\bibitem [{\citenamefont {Navarro}\ \emph {et~al.}(1997)\citenamefont {Navarro}, \citenamefont {Frenk},\ and\ \citenamefont {White}}]{Navarro:1996gj}%
  \BibitemOpen
  \bibfield  {author} {\bibinfo {author} {\bibfnamefont {J.~F.}\ \bibnamefont {Navarro}}, \bibinfo {author} {\bibfnamefont {C.~S.}\ \bibnamefont {Frenk}}, \ and\ \bibinfo {author} {\bibfnamefont {S.~D.~M.}\ \bibnamefont {White}},\ }\href {\doibase 10.1086/304888} {\bibfield  {journal} {\bibinfo  {journal} {Astrophys. J.}\ }\textbf {\bibinfo {volume} {490}},\ \bibinfo {pages} {493} (\bibinfo {year} {1997})},\ \Eprint {http://arxiv.org/abs/astro-ph/9611107} {arXiv:astro-ph/9611107} \BibitemShut {NoStop}%
\bibitem [{\citenamefont {Karukes}\ \emph {et~al.}(2020)\citenamefont {Karukes}, \citenamefont {Benito}, \citenamefont {Iocco}, \citenamefont {Trotta},\ and\ \citenamefont {Geringer-Sameth}}]{Karukes:2019jwa}%
  \BibitemOpen
  \bibfield  {author} {\bibinfo {author} {\bibfnamefont {E.~V.}\ \bibnamefont {Karukes}}, \bibinfo {author} {\bibfnamefont {M.}~\bibnamefont {Benito}}, \bibinfo {author} {\bibfnamefont {F.}~\bibnamefont {Iocco}}, \bibinfo {author} {\bibfnamefont {R.}~\bibnamefont {Trotta}}, \ and\ \bibinfo {author} {\bibfnamefont {A.}~\bibnamefont {Geringer-Sameth}},\ }\href {\doibase 10.1088/1475-7516/2020/05/033} {\bibfield  {journal} {\bibinfo  {journal} {JCAP}\ }\textbf {\bibinfo {volume} {05}},\ \bibinfo {pages} {033} (\bibinfo {year} {2020})},\ \Eprint {http://arxiv.org/abs/1912.04296} {arXiv:1912.04296 [astro-ph.GA]} \BibitemShut {NoStop}%
\bibitem [{\citenamefont {Iguaz}\ \emph {et~al.}(2021)\citenamefont {Iguaz}, \citenamefont {Iguaz}, \citenamefont {Serpico}, \citenamefont {Serpico}, \citenamefont {Siegert},\ and\ \citenamefont {Siegert}}]{Iguaz:2021irx}%
  \BibitemOpen
  \bibfield  {author} {\bibinfo {author} {\bibfnamefont {J.}~\bibnamefont {Iguaz}}, \bibinfo {author} {\bibfnamefont {J.}~\bibnamefont {Iguaz}}, \bibinfo {author} {\bibfnamefont {P.~D.}\ \bibnamefont {Serpico}}, \bibinfo {author} {\bibfnamefont {P.~D.}\ \bibnamefont {Serpico}}, \bibinfo {author} {\bibfnamefont {T.}~\bibnamefont {Siegert}}, \ and\ \bibinfo {author} {\bibfnamefont {T.}~\bibnamefont {Siegert}},\ }\href {\doibase 10.1103/PhysRevD.103.103025} {\bibfield  {journal} {\bibinfo  {journal} {Phys. Rev. D}\ }\textbf {\bibinfo {volume} {103}},\ \bibinfo {pages} {103025} (\bibinfo {year} {2021})},\ \bibinfo {note} {[Erratum: Phys.Rev.D 107, 069902 (2023)]},\ \Eprint {http://arxiv.org/abs/2104.03145} {arXiv:2104.03145 [astro-ph.CO]} \BibitemShut {NoStop}%
\bibitem [{\citenamefont {Essig}\ \emph {et~al.}(2013{\natexlab{a}})\citenamefont {Essig}, \citenamefont {Kuflik}, \citenamefont {McDermott}, \citenamefont {Volansky},\ and\ \citenamefont {Zurek}}]{Essig:2013goa}%
  \BibitemOpen
  \bibfield  {author} {\bibinfo {author} {\bibfnamefont {R.}~\bibnamefont {Essig}}, \bibinfo {author} {\bibfnamefont {E.}~\bibnamefont {Kuflik}}, \bibinfo {author} {\bibfnamefont {S.~D.}\ \bibnamefont {McDermott}}, \bibinfo {author} {\bibfnamefont {T.}~\bibnamefont {Volansky}}, \ and\ \bibinfo {author} {\bibfnamefont {K.~M.}\ \bibnamefont {Zurek}},\ }\href {\doibase 10.1007/JHEP11(2013)193} {\bibfield  {journal} {\bibinfo  {journal} {JHEP}\ }\textbf {\bibinfo {volume} {11}},\ \bibinfo {pages} {193} (\bibinfo {year} {2013}{\natexlab{a}})},\ \Eprint {http://arxiv.org/abs/1309.4091} {arXiv:1309.4091 [hep-ph]} \BibitemShut {NoStop}%
\bibitem [{\citenamefont {Essig}\ \emph {et~al.}(2013{\natexlab{b}})\citenamefont {Essig} \emph {et~al.}}]{Essig:2013lka}%
  \BibitemOpen
  \bibfield  {author} {\bibinfo {author} {\bibfnamefont {R.}~\bibnamefont {Essig}} \emph {et~al.},\ }in\ \href@noop {} {\emph {\bibinfo {booktitle} {{Snowmass 2013}: {Snowmass on the Mississippi}}}}\ (\bibinfo {year} {2013})\ \Eprint {http://arxiv.org/abs/1311.0029} {arXiv:1311.0029 [hep-ph]} \BibitemShut {NoStop}%
\bibitem [{\citenamefont {Ellis}\ \emph {et~al.}(1992)\citenamefont {Ellis}, \citenamefont {Gelmini}, \citenamefont {Lopez}, \citenamefont {Nanopoulos},\ and\ \citenamefont {Sarkar}}]{Ellis:1990nb}%
  \BibitemOpen
  \bibfield  {author} {\bibinfo {author} {\bibfnamefont {J.~R.}\ \bibnamefont {Ellis}}, \bibinfo {author} {\bibfnamefont {G.~B.}\ \bibnamefont {Gelmini}}, \bibinfo {author} {\bibfnamefont {J.~L.}\ \bibnamefont {Lopez}}, \bibinfo {author} {\bibfnamefont {D.~V.}\ \bibnamefont {Nanopoulos}}, \ and\ \bibinfo {author} {\bibfnamefont {S.}~\bibnamefont {Sarkar}},\ }\href {\doibase 10.1016/0550-3213(92)90438-H} {\bibfield  {journal} {\bibinfo  {journal} {Nucl. Phys. B}\ }\textbf {\bibinfo {volume} {373}},\ \bibinfo {pages} {399} (\bibinfo {year} {1992})}\BibitemShut {NoStop}%
\bibitem [{\citenamefont {Iocco}\ \emph {et~al.}(2009)\citenamefont {Iocco}, \citenamefont {Mangano}, \citenamefont {Miele}, \citenamefont {Pisanti},\ and\ \citenamefont {Serpico}}]{Iocco:2008va}%
  \BibitemOpen
  \bibfield  {author} {\bibinfo {author} {\bibfnamefont {F.}~\bibnamefont {Iocco}}, \bibinfo {author} {\bibfnamefont {G.}~\bibnamefont {Mangano}}, \bibinfo {author} {\bibfnamefont {G.}~\bibnamefont {Miele}}, \bibinfo {author} {\bibfnamefont {O.}~\bibnamefont {Pisanti}}, \ and\ \bibinfo {author} {\bibfnamefont {P.~D.}\ \bibnamefont {Serpico}},\ }\href {\doibase 10.1016/j.physrep.2009.02.002} {\bibfield  {journal} {\bibinfo  {journal} {Phys. Rept.}\ }\textbf {\bibinfo {volume} {472}},\ \bibinfo {pages} {1} (\bibinfo {year} {2009})},\ \Eprint {http://arxiv.org/abs/0809.0631} {arXiv:0809.0631 [astro-ph]} \BibitemShut {NoStop}%
\bibitem [{\citenamefont {Holtmann}\ \emph {et~al.}(1999)\citenamefont {Holtmann}, \citenamefont {Kawasaki}, \citenamefont {Kohri},\ and\ \citenamefont {Moroi}}]{Holtmann:1998gd}%
  \BibitemOpen
  \bibfield  {author} {\bibinfo {author} {\bibfnamefont {E.}~\bibnamefont {Holtmann}}, \bibinfo {author} {\bibfnamefont {M.}~\bibnamefont {Kawasaki}}, \bibinfo {author} {\bibfnamefont {K.}~\bibnamefont {Kohri}}, \ and\ \bibinfo {author} {\bibfnamefont {T.}~\bibnamefont {Moroi}},\ }\href {\doibase 10.1103/PhysRevD.60.023506} {\bibfield  {journal} {\bibinfo  {journal} {Phys. Rev. D}\ }\textbf {\bibinfo {volume} {60}},\ \bibinfo {pages} {023506} (\bibinfo {year} {1999})},\ \Eprint {http://arxiv.org/abs/hep-ph/9805405} {arXiv:hep-ph/9805405} \BibitemShut {NoStop}%
\bibitem [{\citenamefont {Kawasaki}\ \emph {et~al.}(2001)\citenamefont {Kawasaki}, \citenamefont {Kohri},\ and\ \citenamefont {Moroi}}]{Kawasaki:2000qr}%
  \BibitemOpen
  \bibfield  {author} {\bibinfo {author} {\bibfnamefont {M.}~\bibnamefont {Kawasaki}}, \bibinfo {author} {\bibfnamefont {K.}~\bibnamefont {Kohri}}, \ and\ \bibinfo {author} {\bibfnamefont {T.}~\bibnamefont {Moroi}},\ }\href {\doibase 10.1103/PhysRevD.63.103502} {\bibfield  {journal} {\bibinfo  {journal} {Phys. Rev. D}\ }\textbf {\bibinfo {volume} {63}},\ \bibinfo {pages} {103502} (\bibinfo {year} {2001})},\ \Eprint {http://arxiv.org/abs/hep-ph/0012279} {arXiv:hep-ph/0012279} \BibitemShut {NoStop}%
\bibitem [{\citenamefont {Siegert}\ \emph {et~al.}(2015)\citenamefont {Siegert}, \citenamefont {Diehl}, \citenamefont {Krause},\ and\ \citenamefont {Greiner}}]{Siegert:2015ila}%
  \BibitemOpen
  \bibfield  {author} {\bibinfo {author} {\bibfnamefont {T.}~\bibnamefont {Siegert}}, \bibinfo {author} {\bibfnamefont {R.}~\bibnamefont {Diehl}}, \bibinfo {author} {\bibfnamefont {M.~G.~H.}\ \bibnamefont {Krause}}, \ and\ \bibinfo {author} {\bibfnamefont {J.}~\bibnamefont {Greiner}},\ }\href {\doibase 10.1051/0004-6361/201525877} {\bibfield  {journal} {\bibinfo  {journal} {Astron. Astrophys.}\ }\textbf {\bibinfo {volume} {579}},\ \bibinfo {pages} {A124} (\bibinfo {year} {2015})},\ \Eprint {http://arxiv.org/abs/1505.05999} {arXiv:1505.05999 [astro-ph.HE]} \BibitemShut {NoStop}%
\bibitem [{\citenamefont {Siegert}\ \emph {et~al.}(2016{\natexlab{a}})\citenamefont {Siegert}, \citenamefont {Diehl}, \citenamefont {Khachatryan}, \citenamefont {Krause}, \citenamefont {Guglielmetti}, \citenamefont {Greiner}, \citenamefont {Strong},\ and\ \citenamefont {Zhang}}]{Siegert:2015knp}%
  \BibitemOpen
  \bibfield  {author} {\bibinfo {author} {\bibfnamefont {T.}~\bibnamefont {Siegert}}, \bibinfo {author} {\bibfnamefont {R.}~\bibnamefont {Diehl}}, \bibinfo {author} {\bibfnamefont {G.}~\bibnamefont {Khachatryan}}, \bibinfo {author} {\bibfnamefont {M.~G.~H.}\ \bibnamefont {Krause}}, \bibinfo {author} {\bibfnamefont {F.}~\bibnamefont {Guglielmetti}}, \bibinfo {author} {\bibfnamefont {J.}~\bibnamefont {Greiner}}, \bibinfo {author} {\bibfnamefont {A.~W.}\ \bibnamefont {Strong}}, \ and\ \bibinfo {author} {\bibfnamefont {X.}~\bibnamefont {Zhang}},\ }\href {\doibase 10.1051/0004-6361/201527510} {\bibfield  {journal} {\bibinfo  {journal} {Astron. Astrophys.}\ }\textbf {\bibinfo {volume} {586}},\ \bibinfo {pages} {A84} (\bibinfo {year} {2016}{\natexlab{a}})},\ \Eprint {http://arxiv.org/abs/1512.00325} {arXiv:1512.00325 [astro-ph.HE]} \BibitemShut {NoStop}%
\bibitem [{\citenamefont {Siegert}\ \emph {et~al.}(2016{\natexlab{b}})\citenamefont {Siegert}, \citenamefont {Diehl}, \citenamefont {Greiner}, \citenamefont {Krause}, \citenamefont {Beloborodov}, \citenamefont {Cadolle~Bel}, \citenamefont {Guglielmetti}, \citenamefont {Rodriguez}, \citenamefont {Strong},\ and\ \citenamefont {Zhang}}]{Siegert:2016ymf}%
  \BibitemOpen
  \bibfield  {author} {\bibinfo {author} {\bibfnamefont {T.}~\bibnamefont {Siegert}}, \bibinfo {author} {\bibfnamefont {R.}~\bibnamefont {Diehl}}, \bibinfo {author} {\bibfnamefont {J.}~\bibnamefont {Greiner}}, \bibinfo {author} {\bibfnamefont {M.~G.~H.}\ \bibnamefont {Krause}}, \bibinfo {author} {\bibfnamefont {A.~M.}\ \bibnamefont {Beloborodov}}, \bibinfo {author} {\bibfnamefont {M.}~\bibnamefont {Cadolle~Bel}}, \bibinfo {author} {\bibfnamefont {F.}~\bibnamefont {Guglielmetti}}, \bibinfo {author} {\bibfnamefont {J.}~\bibnamefont {Rodriguez}}, \bibinfo {author} {\bibfnamefont {A.~W.}\ \bibnamefont {Strong}}, \ and\ \bibinfo {author} {\bibfnamefont {X.}~\bibnamefont {Zhang}},\ }\href {\doibase 10.1038/nature16978} {\bibfield  {journal} {\bibinfo  {journal} {Nature}\ }\textbf {\bibinfo {volume} {531}},\ \bibinfo {pages} {341} (\bibinfo {year} {2016}{\natexlab{b}})},\ \Eprint {http://arxiv.org/abs/1603.01169} {arXiv:1603.01169 [astro-ph.HE]} \BibitemShut {NoStop}%
\bibitem [{\citenamefont {Siegert}\ \emph {et~al.}(2016{\natexlab{c}})\citenamefont {Siegert}, \citenamefont {Diehl}, \citenamefont {Vincent}, \citenamefont {Guglielmetti}, \citenamefont {Krause},\ and\ \citenamefont {Boehm}}]{Siegert:2016ijv}%
  \BibitemOpen
  \bibfield  {author} {\bibinfo {author} {\bibfnamefont {T.}~\bibnamefont {Siegert}}, \bibinfo {author} {\bibfnamefont {R.}~\bibnamefont {Diehl}}, \bibinfo {author} {\bibfnamefont {A.~C.}\ \bibnamefont {Vincent}}, \bibinfo {author} {\bibfnamefont {F.}~\bibnamefont {Guglielmetti}}, \bibinfo {author} {\bibfnamefont {M.~G.~H.}\ \bibnamefont {Krause}}, \ and\ \bibinfo {author} {\bibfnamefont {C.}~\bibnamefont {Boehm}},\ }\href {\doibase 10.1051/0004-6361/201629136} {\bibfield  {journal} {\bibinfo  {journal} {Astron. Astrophys.}\ }\textbf {\bibinfo {volume} {595}},\ \bibinfo {pages} {A25} (\bibinfo {year} {2016}{\natexlab{c}})},\ \Eprint {http://arxiv.org/abs/1608.00393} {arXiv:1608.00393 [astro-ph.HE]} \BibitemShut {NoStop}%
\bibitem [{\citenamefont {Siegert}(2019)}]{Siegert:2019clp}%
  \BibitemOpen
  \bibfield  {author} {\bibinfo {author} {\bibfnamefont {T.}~\bibnamefont {Siegert}},\ }\href {\doibase 10.1051/0004-6361/201936659} {\bibfield  {journal} {\bibinfo  {journal} {Astron. Astrophys.}\ }\textbf {\bibinfo {volume} {632}},\ \bibinfo {pages} {L1} (\bibinfo {year} {2019})},\ \Eprint {http://arxiv.org/abs/1910.09575} {arXiv:1910.09575 [astro-ph.HE]} \BibitemShut {NoStop}%
\bibitem [{\citenamefont {Siegert}\ \emph {et~al.}(2021{\natexlab{a}})\citenamefont {Siegert}, \citenamefont {Crocker}, \citenamefont {Macias}, \citenamefont {Panther}, \citenamefont {Calore}, \citenamefont {Song},\ and\ \citenamefont {Horiuchi}}]{Siegert:2021trw}%
  \BibitemOpen
  \bibfield  {author} {\bibinfo {author} {\bibfnamefont {T.}~\bibnamefont {Siegert}}, \bibinfo {author} {\bibfnamefont {R.~M.}\ \bibnamefont {Crocker}}, \bibinfo {author} {\bibfnamefont {O.}~\bibnamefont {Macias}}, \bibinfo {author} {\bibfnamefont {F.~H.}\ \bibnamefont {Panther}}, \bibinfo {author} {\bibfnamefont {F.}~\bibnamefont {Calore}}, \bibinfo {author} {\bibfnamefont {D.}~\bibnamefont {Song}}, \ and\ \bibinfo {author} {\bibfnamefont {S.}~\bibnamefont {Horiuchi}},\ }\href {\doibase 10.1093/mnrasl/slab113} {\bibfield  {journal} {\bibinfo  {journal} {Mon. Not. Roy. Astron. Soc.}\ }\textbf {\bibinfo {volume} {509}},\ \bibinfo {pages} {L11} (\bibinfo {year} {2021}{\natexlab{a}})},\ \Eprint {http://arxiv.org/abs/2109.03691} {arXiv:2109.03691 [astro-ph.HE]} \BibitemShut {NoStop}%
\bibitem [{\citenamefont {Calore}\ \emph {et~al.}(2023)\citenamefont {Calore}, \citenamefont {Dekker}, \citenamefont {Serpico},\ and\ \citenamefont {Siegert}}]{Calore:2022pks}%
  \BibitemOpen
  \bibfield  {author} {\bibinfo {author} {\bibfnamefont {F.}~\bibnamefont {Calore}}, \bibinfo {author} {\bibfnamefont {A.}~\bibnamefont {Dekker}}, \bibinfo {author} {\bibfnamefont {P.~D.}\ \bibnamefont {Serpico}}, \ and\ \bibinfo {author} {\bibfnamefont {T.}~\bibnamefont {Siegert}},\ }\href {\doibase 10.1093/mnras/stad457} {\bibfield  {journal} {\bibinfo  {journal} {Mon. Not. Roy. Astron. Soc.}\ }\textbf {\bibinfo {volume} {520}},\ \bibinfo {pages} {4167} (\bibinfo {year} {2023})},\ \Eprint {http://arxiv.org/abs/2209.06299} {arXiv:2209.06299 [hep-ph]} \BibitemShut {NoStop}%
\bibitem [{\citenamefont {De~la Torre~Luque}\ \emph {et~al.}(2024{\natexlab{a}})\citenamefont {De~la Torre~Luque}, \citenamefont {Balaji}, \citenamefont {Carenza},\ and\ \citenamefont {Mastrototaro}}]{DelaTorreLuque:2024zsr}%
  \BibitemOpen
  \bibfield  {author} {\bibinfo {author} {\bibfnamefont {P.}~\bibnamefont {De~la Torre~Luque}}, \bibinfo {author} {\bibfnamefont {S.}~\bibnamefont {Balaji}}, \bibinfo {author} {\bibfnamefont {P.}~\bibnamefont {Carenza}}, \ and\ \bibinfo {author} {\bibfnamefont {L.}~\bibnamefont {Mastrototaro}},\ }\href@noop {} {\  (\bibinfo {year} {2024}{\natexlab{a}})},\ \Eprint {http://arxiv.org/abs/2405.08482} {arXiv:2405.08482 [hep-ph]} \BibitemShut {NoStop}%
\bibitem [{\citenamefont {Berteaud}\ \emph {et~al.}(2022)\citenamefont {Berteaud}, \citenamefont {Calore}, \citenamefont {Iguaz}, \citenamefont {Serpico},\ and\ \citenamefont {Siegert}}]{Berteaud:2022tws}%
  \BibitemOpen
  \bibfield  {author} {\bibinfo {author} {\bibfnamefont {J.}~\bibnamefont {Berteaud}}, \bibinfo {author} {\bibfnamefont {F.}~\bibnamefont {Calore}}, \bibinfo {author} {\bibfnamefont {J.}~\bibnamefont {Iguaz}}, \bibinfo {author} {\bibfnamefont {P.~D.}\ \bibnamefont {Serpico}}, \ and\ \bibinfo {author} {\bibfnamefont {T.}~\bibnamefont {Siegert}},\ }\href {\doibase 10.1103/PhysRevD.106.023030} {\bibfield  {journal} {\bibinfo  {journal} {Phys. Rev. D}\ }\textbf {\bibinfo {volume} {106}},\ \bibinfo {pages} {023030} (\bibinfo {year} {2022})},\ \Eprint {http://arxiv.org/abs/2202.07483} {arXiv:2202.07483 [astro-ph.HE]} \BibitemShut {NoStop}%
\bibitem [{\citenamefont {Siegert}\ \emph {et~al.}(2024)\citenamefont {Siegert}, \citenamefont {Calore},\ and\ \citenamefont {Serpico}}]{Siegert:2024hmr}%
  \BibitemOpen
  \bibfield  {author} {\bibinfo {author} {\bibfnamefont {T.}~\bibnamefont {Siegert}}, \bibinfo {author} {\bibfnamefont {F.}~\bibnamefont {Calore}}, \ and\ \bibinfo {author} {\bibfnamefont {P.~D.}\ \bibnamefont {Serpico}},\ }\href {\doibase 10.1093/mnras/stae104} {\bibfield  {journal} {\bibinfo  {journal} {Mon. Not. Roy. Astron. Soc.}\ }\textbf {\bibinfo {volume} {528}},\ \bibinfo {pages} {3433} (\bibinfo {year} {2024})},\ \Eprint {http://arxiv.org/abs/2401.03795} {arXiv:2401.03795 [astro-ph.HE]} \BibitemShut {NoStop}%
\bibitem [{\citenamefont {Siegert}\ \emph {et~al.}(2022)\citenamefont {Siegert}, \citenamefont {Boehm}, \citenamefont {Calore}, \citenamefont {Diehl}, \citenamefont {Krause}, \citenamefont {Serpico},\ and\ \citenamefont {Vincent}}]{Siegert:2021upf}%
  \BibitemOpen
  \bibfield  {author} {\bibinfo {author} {\bibfnamefont {T.}~\bibnamefont {Siegert}}, \bibinfo {author} {\bibfnamefont {C.}~\bibnamefont {Boehm}}, \bibinfo {author} {\bibfnamefont {F.}~\bibnamefont {Calore}}, \bibinfo {author} {\bibfnamefont {R.}~\bibnamefont {Diehl}}, \bibinfo {author} {\bibfnamefont {M.~G.~H.}\ \bibnamefont {Krause}}, \bibinfo {author} {\bibfnamefont {P.~D.}\ \bibnamefont {Serpico}}, \ and\ \bibinfo {author} {\bibfnamefont {A.~C.}\ \bibnamefont {Vincent}},\ }\href {\doibase 10.1093/mnras/stac008} {\bibfield  {journal} {\bibinfo  {journal} {Mon. Not. Roy. Astron. Soc.}\ }\textbf {\bibinfo {volume} {511}},\ \bibinfo {pages} {914} (\bibinfo {year} {2022})},\ \Eprint {http://arxiv.org/abs/2109.03791} {arXiv:2109.03791 [astro-ph.HE]} \BibitemShut {NoStop}%
\bibitem [{\citenamefont {Siegert}\ \emph {et~al.}(2019)\citenamefont {Siegert}, \citenamefont {Diehl}, \citenamefont {Weinberger}, \citenamefont {Pleintinger}, \citenamefont {Greiner},\ and\ \citenamefont {Zhang}}]{Siegert:2019cvk}%
  \BibitemOpen
  \bibfield  {author} {\bibinfo {author} {\bibfnamefont {T.}~\bibnamefont {Siegert}}, \bibinfo {author} {\bibfnamefont {R.}~\bibnamefont {Diehl}}, \bibinfo {author} {\bibfnamefont {C.}~\bibnamefont {Weinberger}}, \bibinfo {author} {\bibfnamefont {M.~M.~M.}\ \bibnamefont {Pleintinger}}, \bibinfo {author} {\bibfnamefont {J.}~\bibnamefont {Greiner}}, \ and\ \bibinfo {author} {\bibfnamefont {X.}~\bibnamefont {Zhang}},\ }\href {\doibase 10.1051/0004-6361/201834920} {\bibfield  {journal} {\bibinfo  {journal} {Astron. Astrophys.}\ }\textbf {\bibinfo {volume} {626}},\ \bibinfo {pages} {A73} (\bibinfo {year} {2019})},\ \Eprint {http://arxiv.org/abs/1903.01096} {arXiv:1903.01096 [astro-ph.HE]} \BibitemShut {NoStop}%
\bibitem [{\citenamefont {{Krivonos, R.}}\ \emph {et~al.}(2007)\citenamefont {{Krivonos, R.}}, \citenamefont {{Revnivtsev, M.}}, \citenamefont {{Churazov, E.}}, \citenamefont {{Sazonov, S.}}, \citenamefont {{Grebenev, S.}},\ and\ \citenamefont {{Sunyaev, R.}}}]{unresolve}%
  \BibitemOpen
  \bibfield  {author} {\bibinfo {author} {\bibnamefont {{Krivonos, R.}}}, \bibinfo {author} {\bibnamefont {{Revnivtsev, M.}}}, \bibinfo {author} {\bibnamefont {{Churazov, E.}}}, \bibinfo {author} {\bibnamefont {{Sazonov, S.}}}, \bibinfo {author} {\bibnamefont {{Grebenev, S.}}}, \ and\ \bibinfo {author} {\bibnamefont {{Sunyaev, R.}}},\ }\href {\doibase 10.1051/0004-6361:20065626} {\bibfield  {journal} {\bibinfo  {journal} {A\&A}\ }\textbf {\bibinfo {volume} {463}},\ \bibinfo {pages} {957} (\bibinfo {year} {2007})}\BibitemShut {NoStop}%
\bibitem [{\citenamefont {Wang}\ \emph {et~al.}(2020)\citenamefont {Wang}, \citenamefont {Siegert}, \citenamefont {Dai}, \citenamefont {Diehl}, \citenamefont {Greiner}, \citenamefont {Heger}, \citenamefont {Krause}, \citenamefont {Lang}, \citenamefont {Pleintinger},\ and\ \citenamefont {Zhang}}]{Wang_2020}%
  \BibitemOpen
  \bibfield  {author} {\bibinfo {author} {\bibfnamefont {W.}~\bibnamefont {Wang}}, \bibinfo {author} {\bibfnamefont {T.}~\bibnamefont {Siegert}}, \bibinfo {author} {\bibfnamefont {Z.~G.}\ \bibnamefont {Dai}}, \bibinfo {author} {\bibfnamefont {R.}~\bibnamefont {Diehl}}, \bibinfo {author} {\bibfnamefont {J.}~\bibnamefont {Greiner}}, \bibinfo {author} {\bibfnamefont {A.}~\bibnamefont {Heger}}, \bibinfo {author} {\bibfnamefont {M.}~\bibnamefont {Krause}}, \bibinfo {author} {\bibfnamefont {M.}~\bibnamefont {Lang}}, \bibinfo {author} {\bibfnamefont {M.~M.~M.}\ \bibnamefont {Pleintinger}}, \ and\ \bibinfo {author} {\bibfnamefont {X.~L.}\ \bibnamefont {Zhang}},\ }\href {\doibase 10.3847/1538-4357/ab6336} {\bibfield  {journal} {\bibinfo  {journal} {The Astrophysical Journal}\ }\textbf {\bibinfo {volume} {889}},\ \bibinfo {pages} {169} (\bibinfo {year} {2020})}\BibitemShut {NoStop}%
\bibitem [{\citenamefont {Siegert}\ \emph {et~al.}(2021{\natexlab{b}})\citenamefont {Siegert}, \citenamefont {Ghosh}, \citenamefont {Mathur}, \citenamefont {Spraggon},\ and\ \citenamefont {Yeddanapudi}}]{Siegert:2021wlq}%
  \BibitemOpen
  \bibfield  {author} {\bibinfo {author} {\bibfnamefont {T.}~\bibnamefont {Siegert}}, \bibinfo {author} {\bibfnamefont {S.}~\bibnamefont {Ghosh}}, \bibinfo {author} {\bibfnamefont {K.}~\bibnamefont {Mathur}}, \bibinfo {author} {\bibfnamefont {E.}~\bibnamefont {Spraggon}}, \ and\ \bibinfo {author} {\bibfnamefont {A.}~\bibnamefont {Yeddanapudi}},\ }\href {\doibase 10.1051/0004-6361/202140300} {\bibfield  {journal} {\bibinfo  {journal} {Astron. Astrophys.}\ }\textbf {\bibinfo {volume} {650}},\ \bibinfo {pages} {A187} (\bibinfo {year} {2021}{\natexlab{b}})},\ \Eprint {http://arxiv.org/abs/2104.00363} {arXiv:2104.00363 [astro-ph.HE]} \BibitemShut {NoStop}%
\bibitem [{\citenamefont {Vianello}\ \emph {et~al.}(2015)\citenamefont {Vianello}, \citenamefont {Lauer}, \citenamefont {Younk}, \citenamefont {Tibaldo}, \citenamefont {Burgess}, \citenamefont {Ayala}, \citenamefont {Harding}, \citenamefont {Hui}, \citenamefont {Omodei},\ and\ \citenamefont {Zhou}}]{Vianello:2015wwa}%
  \BibitemOpen
  \bibfield  {author} {\bibinfo {author} {\bibfnamefont {G.}~\bibnamefont {Vianello}}, \bibinfo {author} {\bibfnamefont {R.~J.}\ \bibnamefont {Lauer}}, \bibinfo {author} {\bibfnamefont {P.}~\bibnamefont {Younk}}, \bibinfo {author} {\bibfnamefont {L.}~\bibnamefont {Tibaldo}}, \bibinfo {author} {\bibfnamefont {J.~M.}\ \bibnamefont {Burgess}}, \bibinfo {author} {\bibfnamefont {H.}~\bibnamefont {Ayala}}, \bibinfo {author} {\bibfnamefont {P.}~\bibnamefont {Harding}}, \bibinfo {author} {\bibfnamefont {M.}~\bibnamefont {Hui}}, \bibinfo {author} {\bibfnamefont {N.}~\bibnamefont {Omodei}}, \ and\ \bibinfo {author} {\bibfnamefont {H.}~\bibnamefont {Zhou}}\ }(\bibinfo {year} {2015})\ \Eprint {http://arxiv.org/abs/1507.08343} {arXiv:1507.08343 [astro-ph.HE]} \BibitemShut {NoStop}%
\bibitem [{\citenamefont {Foreman-Mackey}\ \emph {et~al.}(2013)\citenamefont {Foreman-Mackey}, \citenamefont {Hogg}, \citenamefont {Lang},\ and\ \citenamefont {Goodman}}]{Foreman_Mackey_2013}%
  \BibitemOpen
  \bibfield  {author} {\bibinfo {author} {\bibfnamefont {D.}~\bibnamefont {Foreman-Mackey}}, \bibinfo {author} {\bibfnamefont {D.~W.}\ \bibnamefont {Hogg}}, \bibinfo {author} {\bibfnamefont {D.}~\bibnamefont {Lang}}, \ and\ \bibinfo {author} {\bibfnamefont {J.}~\bibnamefont {Goodman}},\ }\href {\doibase 10.1086/670067} {\bibfield  {journal} {\bibinfo  {journal} {Publications of the Astronomical Society of the Pacific}\ }\textbf {\bibinfo {volume} {125}},\ \bibinfo {pages} {306–312} (\bibinfo {year} {2013})}\BibitemShut {NoStop}%
\bibitem [{\citenamefont {An}\ \emph {et~al.}(2013)\citenamefont {An}, \citenamefont {Pospelov},\ and\ \citenamefont {Pradler}}]{An:2013yfc}%
  \BibitemOpen
  \bibfield  {author} {\bibinfo {author} {\bibfnamefont {H.}~\bibnamefont {An}}, \bibinfo {author} {\bibfnamefont {M.}~\bibnamefont {Pospelov}}, \ and\ \bibinfo {author} {\bibfnamefont {J.}~\bibnamefont {Pradler}},\ }\href {\doibase 10.1016/j.physletb.2013.07.008} {\bibfield  {journal} {\bibinfo  {journal} {Phys. Lett. B}\ }\textbf {\bibinfo {volume} {725}},\ \bibinfo {pages} {190} (\bibinfo {year} {2013})},\ \Eprint {http://arxiv.org/abs/1302.3884} {arXiv:1302.3884 [hep-ph]} \BibitemShut {NoStop}%
\bibitem [{\citenamefont {Churazov}\ \emph {et~al.}(2007)\citenamefont {Churazov} \emph {et~al.}}]{Churazov:2006bk}%
  \BibitemOpen
  \bibfield  {author} {\bibinfo {author} {\bibfnamefont {E.}~\bibnamefont {Churazov}} \emph {et~al.},\ }\href {\doibase 10.1051/0004-6361:20066230} {\bibfield  {journal} {\bibinfo  {journal} {Astron. Astrophys.}\ }\textbf {\bibinfo {volume} {467}},\ \bibinfo {pages} {529} (\bibinfo {year} {2007})},\ \Eprint {http://arxiv.org/abs/astro-ph/0608250} {arXiv:astro-ph/0608250} \BibitemShut {NoStop}%
\bibitem [{\citenamefont {Weidenspointner}\ \emph {et~al.}(2000)\citenamefont {Weidenspointner}, \citenamefont {Varendorff}, \citenamefont {Kappadath}, \citenamefont {Bennett}, \citenamefont {Bloemen}, \citenamefont {Diehl}, \citenamefont {Hermsen}, \citenamefont {Lichti}, \citenamefont {Ryan},\ and\ \citenamefont {Schönfelder}}]{comptel}%
  \BibitemOpen
  \bibfield  {author} {\bibinfo {author} {\bibfnamefont {G.}~\bibnamefont {Weidenspointner}}, \bibinfo {author} {\bibfnamefont {M.}~\bibnamefont {Varendorff}}, \bibinfo {author} {\bibfnamefont {S.~C.}\ \bibnamefont {Kappadath}}, \bibinfo {author} {\bibfnamefont {K.}~\bibnamefont {Bennett}}, \bibinfo {author} {\bibfnamefont {H.}~\bibnamefont {Bloemen}}, \bibinfo {author} {\bibfnamefont {R.}~\bibnamefont {Diehl}}, \bibinfo {author} {\bibfnamefont {W.}~\bibnamefont {Hermsen}}, \bibinfo {author} {\bibfnamefont {G.~G.}\ \bibnamefont {Lichti}}, \bibinfo {author} {\bibfnamefont {J.}~\bibnamefont {Ryan}}, \ and\ \bibinfo {author} {\bibfnamefont {V.}~\bibnamefont {Schönfelder}},\ }\href {\doibase 10.1063/1.1307028} {\bibfield  {journal} {\bibinfo  {journal} {AIP Conference Proceedings}\ }\textbf {\bibinfo {volume} {510}},\ \bibinfo {pages} {467} (\bibinfo {year} {2000})},\ \Eprint {http://arxiv.org/abs/https://pubs.aip.org/aip/acp/article-pdf/510/1/467/11675929/467\_1\_online.pdf}
  {https://pubs.aip.org/aip/acp/article-pdf/510/1/467/11675929/467\_1\_online.pdf} \BibitemShut {NoStop}%
\bibitem [{\citenamefont {Strong}\ \emph {et~al.}(2004)\citenamefont {Strong}, \citenamefont {Moskalenko},\ and\ \citenamefont {Reimer}}]{Strong:2004de}%
  \BibitemOpen
  \bibfield  {author} {\bibinfo {author} {\bibfnamefont {A.~W.}\ \bibnamefont {Strong}}, \bibinfo {author} {\bibfnamefont {I.~V.}\ \bibnamefont {Moskalenko}}, \ and\ \bibinfo {author} {\bibfnamefont {O.}~\bibnamefont {Reimer}},\ }\href {\doibase 10.1086/423193} {\bibfield  {journal} {\bibinfo  {journal} {Astrophys. J.}\ }\textbf {\bibinfo {volume} {613}},\ \bibinfo {pages} {962} (\bibinfo {year} {2004})},\ \Eprint {http://arxiv.org/abs/astro-ph/0406254} {arXiv:astro-ph/0406254} \BibitemShut {NoStop}%
\bibitem [{\citenamefont {Nguyen}\ \emph {et~al.}(2024)\citenamefont {Nguyen}, \citenamefont {John}, \citenamefont {Linden},\ and\ \citenamefont {Tait}}]{Nguyen:2024kwy}%
  \BibitemOpen
  \bibfield  {author} {\bibinfo {author} {\bibfnamefont {T.~T.~Q.}\ \bibnamefont {Nguyen}}, \bibinfo {author} {\bibfnamefont {I.}~\bibnamefont {John}}, \bibinfo {author} {\bibfnamefont {T.}~\bibnamefont {Linden}}, \ and\ \bibinfo {author} {\bibfnamefont {T.~M.~P.}\ \bibnamefont {Tait}},\ }\href@noop {} {\  (\bibinfo {year} {2024})},\ \Eprint {http://arxiv.org/abs/2412.00180} {arXiv:2412.00180 [hep-ph]} \BibitemShut {NoStop}%
\bibitem [{\citenamefont {Nguyen}(2025)}]{Nguyen:2025eva}%
  \BibitemOpen
  \bibfield  {author} {\bibinfo {author} {\bibfnamefont {T.~T.~Q.}\ \bibnamefont {Nguyen}},\ }\href {\doibase 10.22323/1.474.0066} {\bibfield  {journal} {\bibinfo  {journal} {PoS}\ }\textbf {\bibinfo {volume} {COSMICWISPers2024}},\ \bibinfo {pages} {066} (\bibinfo {year} {2025})},\ \Eprint {http://arxiv.org/abs/2506.12153} {arXiv:2506.12153 [hep-ph]} \BibitemShut {NoStop}%
\bibitem [{\citenamefont {Nguyen}\ \emph {et~al.}(2025{\natexlab{b}})\citenamefont {Nguyen}, \citenamefont {De~la Torre~Luque}, \citenamefont {John}, \citenamefont {Balaji}, \citenamefont {Carenza},\ and\ \citenamefont {Linden}}]{Nguyen:2025tkl}%
  \BibitemOpen
  \bibfield  {author} {\bibinfo {author} {\bibfnamefont {T.~T.~Q.}\ \bibnamefont {Nguyen}}, \bibinfo {author} {\bibfnamefont {P.}~\bibnamefont {De~la Torre~Luque}}, \bibinfo {author} {\bibfnamefont {I.}~\bibnamefont {John}}, \bibinfo {author} {\bibfnamefont {S.}~\bibnamefont {Balaji}}, \bibinfo {author} {\bibfnamefont {P.}~\bibnamefont {Carenza}}, \ and\ \bibinfo {author} {\bibfnamefont {T.}~\bibnamefont {Linden}},\ }\href@noop {} {\  (\bibinfo {year} {2025}{\natexlab{b}})},\ \Eprint {http://arxiv.org/abs/2507.13432} {arXiv:2507.13432 [hep-ph]} \BibitemShut {NoStop}%
\bibitem [{\citenamefont {O'Hare}(2020)}]{AxionLimits}%
  \BibitemOpen
  \bibfield  {author} {\bibinfo {author} {\bibfnamefont {C.}~\bibnamefont {O'Hare}},\ }\href {\doibase 10.5281/zenodo.3932430} {\enquote {\bibinfo {title} {cajohare/axionlimits: Axionlimits},}\ }\bibinfo {howpublished} {\url{https://cajohare.github.io/AxionLimits/}} (\bibinfo {year} {2020})\BibitemShut {NoStop}%
\bibitem [{\citenamefont {Wouters}\ and\ \citenamefont {Brun}(2013)}]{Wouters:2013hua}%
  \BibitemOpen
  \bibfield  {author} {\bibinfo {author} {\bibfnamefont {D.}~\bibnamefont {Wouters}}\ and\ \bibinfo {author} {\bibfnamefont {P.}~\bibnamefont {Brun}},\ }\href {\doibase 10.1088/0004-637X/772/1/44} {\bibfield  {journal} {\bibinfo  {journal} {Astrophys. J.}\ }\textbf {\bibinfo {volume} {772}},\ \bibinfo {pages} {44} (\bibinfo {year} {2013})},\ \Eprint {http://arxiv.org/abs/1304.0989} {arXiv:1304.0989 [astro-ph.HE]} \BibitemShut {NoStop}%
\bibitem [{\citenamefont {Marsh}\ \emph {et~al.}(2017)\citenamefont {Marsh}, \citenamefont {Russell}, \citenamefont {Fabian}, \citenamefont {McNamara}, \citenamefont {Nulsen},\ and\ \citenamefont {Reynolds}}]{Marsh:2017yvc}%
  \BibitemOpen
  \bibfield  {author} {\bibinfo {author} {\bibfnamefont {M.~C.~D.}\ \bibnamefont {Marsh}}, \bibinfo {author} {\bibfnamefont {H.~R.}\ \bibnamefont {Russell}}, \bibinfo {author} {\bibfnamefont {A.~C.}\ \bibnamefont {Fabian}}, \bibinfo {author} {\bibfnamefont {B.~P.}\ \bibnamefont {McNamara}}, \bibinfo {author} {\bibfnamefont {P.}~\bibnamefont {Nulsen}}, \ and\ \bibinfo {author} {\bibfnamefont {C.~S.}\ \bibnamefont {Reynolds}},\ }\href {\doibase 10.1088/1475-7516/2017/12/036} {\bibfield  {journal} {\bibinfo  {journal} {JCAP}\ }\textbf {\bibinfo {volume} {12}},\ \bibinfo {pages} {036} (\bibinfo {year} {2017})},\ \Eprint {http://arxiv.org/abs/1703.07354} {arXiv:1703.07354 [hep-ph]} \BibitemShut {NoStop}%
\bibitem [{\citenamefont {Reynolds}\ \emph {et~al.}(2020)\citenamefont {Reynolds}, \citenamefont {Marsh}, \citenamefont {Russell}, \citenamefont {Fabian}, \citenamefont {Smith}, \citenamefont {Tombesi},\ and\ \citenamefont {Veilleux}}]{Reynolds:2019uqt}%
  \BibitemOpen
  \bibfield  {author} {\bibinfo {author} {\bibfnamefont {C.~S.}\ \bibnamefont {Reynolds}}, \bibinfo {author} {\bibfnamefont {M.~C.~D.}\ \bibnamefont {Marsh}}, \bibinfo {author} {\bibfnamefont {H.~R.}\ \bibnamefont {Russell}}, \bibinfo {author} {\bibfnamefont {A.~C.}\ \bibnamefont {Fabian}}, \bibinfo {author} {\bibfnamefont {R.}~\bibnamefont {Smith}}, \bibinfo {author} {\bibfnamefont {F.}~\bibnamefont {Tombesi}}, \ and\ \bibinfo {author} {\bibfnamefont {S.}~\bibnamefont {Veilleux}},\ }\href {\doibase 10.3847/1538-4357/ab6a0c} {\bibfield  {journal} {\bibinfo  {journal} {Astrophys. J.}\ }\textbf {\bibinfo {volume} {890}},\ \bibinfo {pages} {59} (\bibinfo {year} {2020})},\ \Eprint {http://arxiv.org/abs/1907.05475} {arXiv:1907.05475 [hep-ph]} \BibitemShut {NoStop}%
\bibitem [{\citenamefont {Reyn\'es}\ \emph {et~al.}(2021)\citenamefont {Reyn\'es}, \citenamefont {Matthews}, \citenamefont {Reynolds}, \citenamefont {Russell}, \citenamefont {Smith},\ and\ \citenamefont {Marsh}}]{Reynes:2021bpe}%
  \BibitemOpen
  \bibfield  {author} {\bibinfo {author} {\bibfnamefont {J.~S.}\ \bibnamefont {Reyn\'es}}, \bibinfo {author} {\bibfnamefont {J.~H.}\ \bibnamefont {Matthews}}, \bibinfo {author} {\bibfnamefont {C.~S.}\ \bibnamefont {Reynolds}}, \bibinfo {author} {\bibfnamefont {H.~R.}\ \bibnamefont {Russell}}, \bibinfo {author} {\bibfnamefont {R.~N.}\ \bibnamefont {Smith}}, \ and\ \bibinfo {author} {\bibfnamefont {M.~C.~D.}\ \bibnamefont {Marsh}},\ }\href {\doibase 10.1093/mnras/stab3464} {\bibfield  {journal} {\bibinfo  {journal} {Mon. Not. Roy. Astron. Soc.}\ }\textbf {\bibinfo {volume} {510}},\ \bibinfo {pages} {1264} (\bibinfo {year} {2021})},\ \Eprint {http://arxiv.org/abs/2109.03261} {arXiv:2109.03261 [astro-ph.HE]} \BibitemShut {NoStop}%
\bibitem [{\citenamefont {Foster}\ \emph {et~al.}(2021)\citenamefont {Foster}, \citenamefont {Kongsore}, \citenamefont {Dessert}, \citenamefont {Park}, \citenamefont {Rodd}, \citenamefont {Cranmer},\ and\ \citenamefont {Safdi}}]{Foster:2021ngm}%
  \BibitemOpen
  \bibfield  {author} {\bibinfo {author} {\bibfnamefont {J.~W.}\ \bibnamefont {Foster}}, \bibinfo {author} {\bibfnamefont {M.}~\bibnamefont {Kongsore}}, \bibinfo {author} {\bibfnamefont {C.}~\bibnamefont {Dessert}}, \bibinfo {author} {\bibfnamefont {Y.}~\bibnamefont {Park}}, \bibinfo {author} {\bibfnamefont {N.~L.}\ \bibnamefont {Rodd}}, \bibinfo {author} {\bibfnamefont {K.}~\bibnamefont {Cranmer}}, \ and\ \bibinfo {author} {\bibfnamefont {B.~R.}\ \bibnamefont {Safdi}},\ }\href {\doibase 10.1103/PhysRevLett.127.051101} {\bibfield  {journal} {\bibinfo  {journal} {Phys. Rev. Lett.}\ }\textbf {\bibinfo {volume} {127}},\ \bibinfo {pages} {051101} (\bibinfo {year} {2021})},\ \Eprint {http://arxiv.org/abs/2102.02207} {arXiv:2102.02207 [astro-ph.CO]} \BibitemShut {NoStop}%
\bibitem [{\citenamefont {De~la Torre~Luque}\ \emph {et~al.}(2024{\natexlab{b}})\citenamefont {De~la Torre~Luque}, \citenamefont {Balaji},\ and\ \citenamefont {Carenza}}]{DelaTorreLuque:2023nhh}%
  \BibitemOpen
  \bibfield  {author} {\bibinfo {author} {\bibfnamefont {P.}~\bibnamefont {De~la Torre~Luque}}, \bibinfo {author} {\bibfnamefont {S.}~\bibnamefont {Balaji}}, \ and\ \bibinfo {author} {\bibfnamefont {P.}~\bibnamefont {Carenza}},\ }\href {\doibase 10.1103/PhysRevD.109.L101305} {\bibfield  {journal} {\bibinfo  {journal} {Phys. Rev. D}\ }\textbf {\bibinfo {volume} {109}},\ \bibinfo {pages} {L101305} (\bibinfo {year} {2024}{\natexlab{b}})},\ \Eprint {http://arxiv.org/abs/2307.13728} {arXiv:2307.13728 [hep-ph]} \BibitemShut {NoStop}%
\bibitem [{\citenamefont {Perez}\ \emph {et~al.}(2017)\citenamefont {Perez}, \citenamefont {Ng}, \citenamefont {Beacom}, \citenamefont {Hersh}, \citenamefont {Horiuchi},\ and\ \citenamefont {Krivonos}}]{Perez:2016tcq}%
  \BibitemOpen
  \bibfield  {author} {\bibinfo {author} {\bibfnamefont {K.}~\bibnamefont {Perez}}, \bibinfo {author} {\bibfnamefont {K.~C.~Y.}\ \bibnamefont {Ng}}, \bibinfo {author} {\bibfnamefont {J.~F.}\ \bibnamefont {Beacom}}, \bibinfo {author} {\bibfnamefont {C.}~\bibnamefont {Hersh}}, \bibinfo {author} {\bibfnamefont {S.}~\bibnamefont {Horiuchi}}, \ and\ \bibinfo {author} {\bibfnamefont {R.}~\bibnamefont {Krivonos}},\ }\href {\doibase 10.1103/PhysRevD.95.123002} {\bibfield  {journal} {\bibinfo  {journal} {Phys. Rev. D}\ }\textbf {\bibinfo {volume} {95}},\ \bibinfo {pages} {123002} (\bibinfo {year} {2017})},\ \Eprint {http://arxiv.org/abs/1609.00667} {arXiv:1609.00667 [astro-ph.HE]} \BibitemShut {NoStop}%
\bibitem [{\citenamefont {Ng}\ \emph {et~al.}(2019)\citenamefont {Ng}, \citenamefont {Roach}, \citenamefont {Perez}, \citenamefont {Beacom}, \citenamefont {Horiuchi}, \citenamefont {Krivonos},\ and\ \citenamefont {Wik}}]{Ng:2019gch}%
  \BibitemOpen
  \bibfield  {author} {\bibinfo {author} {\bibfnamefont {K.~C.~Y.}\ \bibnamefont {Ng}}, \bibinfo {author} {\bibfnamefont {B.~M.}\ \bibnamefont {Roach}}, \bibinfo {author} {\bibfnamefont {K.}~\bibnamefont {Perez}}, \bibinfo {author} {\bibfnamefont {J.~F.}\ \bibnamefont {Beacom}}, \bibinfo {author} {\bibfnamefont {S.}~\bibnamefont {Horiuchi}}, \bibinfo {author} {\bibfnamefont {R.}~\bibnamefont {Krivonos}}, \ and\ \bibinfo {author} {\bibfnamefont {D.~R.}\ \bibnamefont {Wik}},\ }\href {\doibase 10.1103/PhysRevD.99.083005} {\bibfield  {journal} {\bibinfo  {journal} {Phys. Rev. D}\ }\textbf {\bibinfo {volume} {99}},\ \bibinfo {pages} {083005} (\bibinfo {year} {2019})},\ \Eprint {http://arxiv.org/abs/1901.01262} {arXiv:1901.01262 [astro-ph.HE]} \BibitemShut {NoStop}%
\bibitem [{\citenamefont {Roach}\ \emph {et~al.}(2023)\citenamefont {Roach}, \citenamefont {Rossland}, \citenamefont {Ng}, \citenamefont {Perez}, \citenamefont {Beacom}, \citenamefont {Grefenstette}, \citenamefont {Horiuchi}, \citenamefont {Krivonos},\ and\ \citenamefont {Wik}}]{Roach:2022lgo}%
  \BibitemOpen
  \bibfield  {author} {\bibinfo {author} {\bibfnamefont {B.~M.}\ \bibnamefont {Roach}}, \bibinfo {author} {\bibfnamefont {S.}~\bibnamefont {Rossland}}, \bibinfo {author} {\bibfnamefont {K.~C.~Y.}\ \bibnamefont {Ng}}, \bibinfo {author} {\bibfnamefont {K.}~\bibnamefont {Perez}}, \bibinfo {author} {\bibfnamefont {J.~F.}\ \bibnamefont {Beacom}}, \bibinfo {author} {\bibfnamefont {B.~W.}\ \bibnamefont {Grefenstette}}, \bibinfo {author} {\bibfnamefont {S.}~\bibnamefont {Horiuchi}}, \bibinfo {author} {\bibfnamefont {R.}~\bibnamefont {Krivonos}}, \ and\ \bibinfo {author} {\bibfnamefont {D.~R.}\ \bibnamefont {Wik}},\ }\href {\doibase 10.1103/PhysRevD.107.023009} {\bibfield  {journal} {\bibinfo  {journal} {Phys. Rev. D}\ }\textbf {\bibinfo {volume} {107}},\ \bibinfo {pages} {023009} (\bibinfo {year} {2023})},\ \Eprint {http://arxiv.org/abs/2207.04572} {arXiv:2207.04572 [astro-ph.HE]} \BibitemShut {NoStop}%
\bibitem [{\citenamefont {Fong}\ \emph {et~al.}(2024)\citenamefont {Fong}, \citenamefont {Ng},\ and\ \citenamefont {Liu}}]{Fong:2024qeq}%
  \BibitemOpen
  \bibfield  {author} {\bibinfo {author} {\bibfnamefont {C.}~\bibnamefont {Fong}}, \bibinfo {author} {\bibfnamefont {K.~C.~Y.}\ \bibnamefont {Ng}}, \ and\ \bibinfo {author} {\bibfnamefont {Q.}~\bibnamefont {Liu}},\ }\href@noop {} {\  (\bibinfo {year} {2024})},\ \Eprint {http://arxiv.org/abs/2401.16747} {arXiv:2401.16747 [hep-ph]} \BibitemShut {NoStop}%
\bibitem [{\citenamefont {Janish}\ and\ \citenamefont {Pinetti}(2023)}]{Janish:2023kvi}%
  \BibitemOpen
  \bibfield  {author} {\bibinfo {author} {\bibfnamefont {R.}~\bibnamefont {Janish}}\ and\ \bibinfo {author} {\bibfnamefont {E.}~\bibnamefont {Pinetti}},\ }\href@noop {} {\  (\bibinfo {year} {2023})},\ \Eprint {http://arxiv.org/abs/2310.15395} {arXiv:2310.15395 [hep-ph]} \BibitemShut {NoStop}%
\bibitem [{\citenamefont {An}\ \emph {et~al.}(2024)\citenamefont {An}, \citenamefont {Ge}, \citenamefont {Liu},\ and\ \citenamefont {Lu}}]{An:2024kls}%
  \BibitemOpen
  \bibfield  {author} {\bibinfo {author} {\bibfnamefont {H.}~\bibnamefont {An}}, \bibinfo {author} {\bibfnamefont {S.}~\bibnamefont {Ge}}, \bibinfo {author} {\bibfnamefont {J.}~\bibnamefont {Liu}}, \ and\ \bibinfo {author} {\bibfnamefont {Z.}~\bibnamefont {Lu}},\ }\href@noop {} {\  (\bibinfo {year} {2024})},\ \Eprint {http://arxiv.org/abs/2402.17140} {arXiv:2402.17140 [hep-ph]} \BibitemShut {NoStop}%
\bibitem [{\citenamefont {Roy}\ \emph {et~al.}(2023)\citenamefont {Roy}, \citenamefont {Blanco}, \citenamefont {Dessert}, \citenamefont {Prabhu},\ and\ \citenamefont {Temim}}]{Roy:2023omw}%
  \BibitemOpen
  \bibfield  {author} {\bibinfo {author} {\bibfnamefont {S.}~\bibnamefont {Roy}}, \bibinfo {author} {\bibfnamefont {C.}~\bibnamefont {Blanco}}, \bibinfo {author} {\bibfnamefont {C.}~\bibnamefont {Dessert}}, \bibinfo {author} {\bibfnamefont {A.}~\bibnamefont {Prabhu}}, \ and\ \bibinfo {author} {\bibfnamefont {T.}~\bibnamefont {Temim}},\ }\href@noop {} {\  (\bibinfo {year} {2023})},\ \Eprint {http://arxiv.org/abs/2311.04987} {arXiv:2311.04987 [hep-ph]} \BibitemShut {NoStop}%
\bibitem [{\citenamefont {Bessho}\ \emph {et~al.}(2022)\citenamefont {Bessho}, \citenamefont {Ikeda},\ and\ \citenamefont {Yin}}]{Bessho:2022yyu}%
  \BibitemOpen
  \bibfield  {author} {\bibinfo {author} {\bibfnamefont {T.}~\bibnamefont {Bessho}}, \bibinfo {author} {\bibfnamefont {Y.}~\bibnamefont {Ikeda}}, \ and\ \bibinfo {author} {\bibfnamefont {W.}~\bibnamefont {Yin}},\ }\href {\doibase 10.1103/PhysRevD.106.095025} {\bibfield  {journal} {\bibinfo  {journal} {Phys. Rev. D}\ }\textbf {\bibinfo {volume} {106}},\ \bibinfo {pages} {095025} (\bibinfo {year} {2022})},\ \Eprint {http://arxiv.org/abs/2208.05975} {arXiv:2208.05975 [hep-ph]} \BibitemShut {NoStop}%
\bibitem [{\citenamefont {Yin}\ \emph {et~al.}(2024)\citenamefont {Yin} \emph {et~al.}}]{Yin:2024lla}%
  \BibitemOpen
  \bibfield  {author} {\bibinfo {author} {\bibfnamefont {W.}~\bibnamefont {Yin}} \emph {et~al.},\ }\href@noop {} {\  (\bibinfo {year} {2024})},\ \Eprint {http://arxiv.org/abs/2402.07976} {arXiv:2402.07976 [astro-ph.CO]} \BibitemShut {NoStop}%
\bibitem [{\citenamefont {Todarello}\ \emph {et~al.}(2024)\citenamefont {Todarello}, \citenamefont {Regis}, \citenamefont {Reynoso-Cordova}, \citenamefont {Taoso}, \citenamefont {Vaz}, \citenamefont {Brinchmann}, \citenamefont {Steinmetz},\ and\ \citenamefont {Zoutendijke}}]{Todarello:2023hdk}%
  \BibitemOpen
  \bibfield  {author} {\bibinfo {author} {\bibfnamefont {E.}~\bibnamefont {Todarello}}, \bibinfo {author} {\bibfnamefont {M.}~\bibnamefont {Regis}}, \bibinfo {author} {\bibfnamefont {J.}~\bibnamefont {Reynoso-Cordova}}, \bibinfo {author} {\bibfnamefont {M.}~\bibnamefont {Taoso}}, \bibinfo {author} {\bibfnamefont {D.}~\bibnamefont {Vaz}}, \bibinfo {author} {\bibfnamefont {J.}~\bibnamefont {Brinchmann}}, \bibinfo {author} {\bibfnamefont {M.}~\bibnamefont {Steinmetz}}, \ and\ \bibinfo {author} {\bibfnamefont {S.~L.}\ \bibnamefont {Zoutendijke}},\ }\href {\doibase 10.1088/1475-7516/2024/05/043} {\bibfield  {journal} {\bibinfo  {journal} {JCAP}\ }\textbf {\bibinfo {volume} {05}},\ \bibinfo {pages} {043} (\bibinfo {year} {2024})},\ \Eprint {http://arxiv.org/abs/2307.07403} {arXiv:2307.07403 [astro-ph.CO]} \BibitemShut {NoStop}%
\bibitem [{\citenamefont {Thorpe-Morgan}\ \emph {et~al.}(2020)\citenamefont {Thorpe-Morgan}, \citenamefont {Malyshev}, \citenamefont {Santangelo}, \citenamefont {Jochum}, \citenamefont {J\"ager}, \citenamefont {Sasaki},\ and\ \citenamefont {Saeedi}}]{Thorpe-Morgan:2020rwc}%
  \BibitemOpen
  \bibfield  {author} {\bibinfo {author} {\bibfnamefont {C.}~\bibnamefont {Thorpe-Morgan}}, \bibinfo {author} {\bibfnamefont {D.}~\bibnamefont {Malyshev}}, \bibinfo {author} {\bibfnamefont {A.}~\bibnamefont {Santangelo}}, \bibinfo {author} {\bibfnamefont {J.}~\bibnamefont {Jochum}}, \bibinfo {author} {\bibfnamefont {B.}~\bibnamefont {J\"ager}}, \bibinfo {author} {\bibfnamefont {M.}~\bibnamefont {Sasaki}}, \ and\ \bibinfo {author} {\bibfnamefont {S.}~\bibnamefont {Saeedi}},\ }\href {\doibase 10.1103/PhysRevD.102.123003} {\bibfield  {journal} {\bibinfo  {journal} {Phys. Rev. D}\ }\textbf {\bibinfo {volume} {102}},\ \bibinfo {pages} {123003} (\bibinfo {year} {2020})},\ \Eprint {http://arxiv.org/abs/2008.08306} {arXiv:2008.08306 [astro-ph.HE]} \BibitemShut {NoStop}%
\bibitem [{\citenamefont {De~Angelis}\ \emph {et~al.}(2021)\citenamefont {De~Angelis} \emph {et~al.}}]{De_Angelis_2021}%
  \BibitemOpen
  \bibfield  {author} {\bibinfo {author} {\bibfnamefont {A.}~\bibnamefont {De~Angelis}} \emph {et~al.},\ }\href {\doibase 10.1007/s10686-021-09706-y} {\bibfield  {journal} {\bibinfo  {journal} {Experimental Astronomy}\ }\textbf {\bibinfo {volume} {51}},\ \bibinfo {pages} {1225–1254} (\bibinfo {year} {2021})}\BibitemShut {NoStop}%
\bibitem [{\citenamefont {Dessert}\ \emph {et~al.}(2024)\citenamefont {Dessert}, \citenamefont {Ning}, \citenamefont {Rodd},\ and\ \citenamefont {Safdi}}]{Dessert:2023vyl}%
  \BibitemOpen
  \bibfield  {author} {\bibinfo {author} {\bibfnamefont {C.}~\bibnamefont {Dessert}}, \bibinfo {author} {\bibfnamefont {O.}~\bibnamefont {Ning}}, \bibinfo {author} {\bibfnamefont {N.~L.}\ \bibnamefont {Rodd}}, \ and\ \bibinfo {author} {\bibfnamefont {B.~R.}\ \bibnamefont {Safdi}},\ }\href {\doibase 10.1103/PhysRevLett.132.211002} {\bibfield  {journal} {\bibinfo  {journal} {Phys. Rev. Lett.}\ }\textbf {\bibinfo {volume} {132}},\ \bibinfo {pages} {211002} (\bibinfo {year} {2024})},\ \Eprint {http://arxiv.org/abs/2305.17160} {arXiv:2305.17160 [astro-ph.CO]} \BibitemShut {NoStop}%
\bibitem [{\citenamefont {Barret}\ \emph {et~al.}(2013)\citenamefont {Barret} \emph {et~al.}}]{Barret:2013mxa}%
  \BibitemOpen
  \bibfield  {author} {\bibinfo {author} {\bibfnamefont {D.}~\bibnamefont {Barret}} \emph {et~al.},\ }\href@noop {} {\  (\bibinfo {year} {2013})},\ \Eprint {http://arxiv.org/abs/1308.6784} {arXiv:1308.6784 [astro-ph.IM]} \BibitemShut {NoStop}%
\bibitem [{\citenamefont {Sisk-Reyn\'es}\ \emph {et~al.}(2023)\citenamefont {Sisk-Reyn\'es}, \citenamefont {Reynolds}, \citenamefont {Parker}, \citenamefont {Matthews},\ and\ \citenamefont {Marsh}}]{Sisk-Reynes:2022sqd}%
  \BibitemOpen
  \bibfield  {author} {\bibinfo {author} {\bibfnamefont {J.}~\bibnamefont {Sisk-Reyn\'es}}, \bibinfo {author} {\bibfnamefont {C.~S.}\ \bibnamefont {Reynolds}}, \bibinfo {author} {\bibfnamefont {M.~L.}\ \bibnamefont {Parker}}, \bibinfo {author} {\bibfnamefont {J.~H.}\ \bibnamefont {Matthews}}, \ and\ \bibinfo {author} {\bibfnamefont {M.~C.~D.}\ \bibnamefont {Marsh}},\ }\href {\doibase 10.3847/1538-4357/acd116} {\bibfield  {journal} {\bibinfo  {journal} {Astrophys. J.}\ }\textbf {\bibinfo {volume} {951}},\ \bibinfo {pages} {5} (\bibinfo {year} {2023})},\ \Eprint {http://arxiv.org/abs/2211.05136} {arXiv:2211.05136 [astro-ph.HE]} \BibitemShut {NoStop}%
\bibitem [{\citenamefont {Halliday}\ \emph {et~al.}(2024)\citenamefont {Halliday} \emph {et~al.}}]{Halliday:2024lca}%
  \BibitemOpen
  \bibfield  {author} {\bibinfo {author} {\bibfnamefont {J.~W.~D.}\ \bibnamefont {Halliday}} \emph {et~al.},\ }\href@noop {} {\  (\bibinfo {year} {2024})},\ \Eprint {http://arxiv.org/abs/2404.17333} {arXiv:2404.17333 [hep-ph]} \BibitemShut {NoStop}%
\bibitem [{\citenamefont {Harris}\ \emph {et~al.}(2020)\citenamefont {Harris}, \citenamefont {Millman}, \citenamefont {van~der Walt}, \citenamefont {Gommers}, \citenamefont {Virtanen}, \citenamefont {Cournapeau}, \citenamefont {Wieser}, \citenamefont {Taylor}, \citenamefont {Berg}, \citenamefont {Smith}, \citenamefont {Kern}, \citenamefont {Picus}, \citenamefont {Hoyer}, \citenamefont {van Kerkwijk}, \citenamefont {Brett}, \citenamefont {Haldane}, \citenamefont {del Río}, \citenamefont {Wiebe}, \citenamefont {Peterson}, \citenamefont {Gérard-Marchant}, \citenamefont {Sheppard}, \citenamefont {Reddy}, \citenamefont {Weckesser}, \citenamefont {Abbasi}, \citenamefont {Gohlke},\ and\ \citenamefont {Oliphant}}]{Harris_2020}%
  \BibitemOpen
  \bibfield  {author} {\bibinfo {author} {\bibfnamefont {C.~R.}\ \bibnamefont {Harris}}, \bibinfo {author} {\bibfnamefont {K.~J.}\ \bibnamefont {Millman}}, \bibinfo {author} {\bibfnamefont {S.~J.}\ \bibnamefont {van~der Walt}}, \bibinfo {author} {\bibfnamefont {R.}~\bibnamefont {Gommers}}, \bibinfo {author} {\bibfnamefont {P.}~\bibnamefont {Virtanen}}, \bibinfo {author} {\bibfnamefont {D.}~\bibnamefont {Cournapeau}}, \bibinfo {author} {\bibfnamefont {E.}~\bibnamefont {Wieser}}, \bibinfo {author} {\bibfnamefont {J.}~\bibnamefont {Taylor}}, \bibinfo {author} {\bibfnamefont {S.}~\bibnamefont {Berg}}, \bibinfo {author} {\bibfnamefont {N.~J.}\ \bibnamefont {Smith}}, \bibinfo {author} {\bibfnamefont {R.}~\bibnamefont {Kern}}, \bibinfo {author} {\bibfnamefont {M.}~\bibnamefont {Picus}}, \bibinfo {author} {\bibfnamefont {S.}~\bibnamefont {Hoyer}}, \bibinfo {author} {\bibfnamefont {M.~H.}\ \bibnamefont {van Kerkwijk}}, \bibinfo {author} {\bibfnamefont {M.}~\bibnamefont {Brett}}, \bibinfo {author} {\bibfnamefont
  {A.}~\bibnamefont {Haldane}}, \bibinfo {author} {\bibfnamefont {J.~F.}\ \bibnamefont {del Río}}, \bibinfo {author} {\bibfnamefont {M.}~\bibnamefont {Wiebe}}, \bibinfo {author} {\bibfnamefont {P.}~\bibnamefont {Peterson}}, \bibinfo {author} {\bibfnamefont {P.}~\bibnamefont {Gérard-Marchant}}, \bibinfo {author} {\bibfnamefont {K.}~\bibnamefont {Sheppard}}, \bibinfo {author} {\bibfnamefont {T.}~\bibnamefont {Reddy}}, \bibinfo {author} {\bibfnamefont {W.}~\bibnamefont {Weckesser}}, \bibinfo {author} {\bibfnamefont {H.}~\bibnamefont {Abbasi}}, \bibinfo {author} {\bibfnamefont {C.}~\bibnamefont {Gohlke}}, \ and\ \bibinfo {author} {\bibfnamefont {T.~E.}\ \bibnamefont {Oliphant}},\ }\href {\doibase 10.1038/s41586-020-2649-2} {\bibfield  {journal} {\bibinfo  {journal} {Nature}\ }\textbf {\bibinfo {volume} {585}},\ \bibinfo {pages} {357–362} (\bibinfo {year} {2020})}\BibitemShut {NoStop}%
\bibitem [{\citenamefont {Virtanen}\ \emph {et~al.}(2020)\citenamefont {Virtanen} \emph {et~al.}}]{Virtanen:2019joe}%
  \BibitemOpen
  \bibfield  {author} {\bibinfo {author} {\bibfnamefont {P.}~\bibnamefont {Virtanen}} \emph {et~al.},\ }\href {\doibase 10.1038/s41592-019-0686-2} {\bibfield  {journal} {\bibinfo  {journal} {Nature Meth.}\ }\textbf {\bibinfo {volume} {17}},\ \bibinfo {pages} {261} (\bibinfo {year} {2020})},\ \Eprint {http://arxiv.org/abs/1907.10121} {arXiv:1907.10121 [cs.MS]} \BibitemShut {NoStop}%
\bibitem [{\citenamefont {Robitaille}\ \emph {et~al.}(2013)\citenamefont {Robitaille} \emph {et~al.}}]{Astropy:2013muo}%
  \BibitemOpen
  \bibfield  {author} {\bibinfo {author} {\bibfnamefont {T.~P.}\ \bibnamefont {Robitaille}} \emph {et~al.} (\bibinfo {collaboration} {Astropy}),\ }\href {\doibase 10.1051/0004-6361/201322068} {\bibfield  {journal} {\bibinfo  {journal} {Astron. Astrophys.}\ }\textbf {\bibinfo {volume} {558}},\ \bibinfo {pages} {A33} (\bibinfo {year} {2013})},\ \Eprint {http://arxiv.org/abs/1307.6212} {arXiv:1307.6212 [astro-ph.IM]} \BibitemShut {NoStop}%
\bibitem [{\citenamefont {Hunter}(2007)}]{HunterMatplotlib}%
  \BibitemOpen
  \bibfield  {author} {\bibinfo {author} {\bibfnamefont {J.~D.}\ \bibnamefont {Hunter}},\ }\href {\doibase 10.1109/MCSE.2007.55} {\bibfield  {journal} {\bibinfo  {journal} {Computing in Science \& Engineering}\ }\textbf {\bibinfo {volume} {9}},\ \bibinfo {pages} {90} (\bibinfo {year} {2007})}\BibitemShut {NoStop}%
\bibitem [{\citenamefont {{Kluyver}}\ \emph {et~al.}(2016)\citenamefont {{Kluyver}}, \citenamefont {{Ragan-Kelley}}, \citenamefont {{P{\'e}rez}}, \citenamefont {{Granger}}, \citenamefont {{Bussonnier}}, \citenamefont {{Frederic}}, \citenamefont {{Kelley}}, \citenamefont {{Hamrick}}, \citenamefont {{Grout}}, \citenamefont {{Corlay}}, \citenamefont {{Ivanov}}, \citenamefont {{Avila}}, \citenamefont {{Abdalla}}, \citenamefont {{Willing}},\ and\ \citenamefont {{Jupyter Development Team}}}]{2016ppap.book...87K}%
  \BibitemOpen
  \bibfield  {author} {\bibinfo {author} {\bibfnamefont {T.}~\bibnamefont {{Kluyver}}}, \bibinfo {author} {\bibfnamefont {B.}~\bibnamefont {{Ragan-Kelley}}}, \bibinfo {author} {\bibfnamefont {F.}~\bibnamefont {{P{\'e}rez}}}, \bibinfo {author} {\bibfnamefont {B.}~\bibnamefont {{Granger}}}, \bibinfo {author} {\bibfnamefont {M.}~\bibnamefont {{Bussonnier}}}, \bibinfo {author} {\bibfnamefont {J.}~\bibnamefont {{Frederic}}}, \bibinfo {author} {\bibfnamefont {K.}~\bibnamefont {{Kelley}}}, \bibinfo {author} {\bibfnamefont {J.}~\bibnamefont {{Hamrick}}}, \bibinfo {author} {\bibfnamefont {J.}~\bibnamefont {{Grout}}}, \bibinfo {author} {\bibfnamefont {S.}~\bibnamefont {{Corlay}}}, \bibinfo {author} {\bibfnamefont {P.}~\bibnamefont {{Ivanov}}}, \bibinfo {author} {\bibfnamefont {D.}~\bibnamefont {{Avila}}}, \bibinfo {author} {\bibfnamefont {S.}~\bibnamefont {{Abdalla}}}, \bibinfo {author} {\bibfnamefont {C.}~\bibnamefont {{Willing}}}, \ and\ \bibinfo {author} {\bibnamefont {{Jupyter Development Team}}},\ }in\ \href
  {\doibase 10.3233/978-1-61499-649-1-87} {\emph {\bibinfo {booktitle} {IOS Press}}}\ (\bibinfo {year} {2016})\ pp.\ \bibinfo {pages} {87--90}\BibitemShut {NoStop}%
\bibitem [{\citenamefont {Binosi}\ \emph {et~al.}(2009)\citenamefont {Binosi}, \citenamefont {Collins}, \citenamefont {Kaufhold},\ and\ \citenamefont {Theussl}}]{Binosi:2008ig}%
  \BibitemOpen
  \bibfield  {author} {\bibinfo {author} {\bibfnamefont {D.}~\bibnamefont {Binosi}}, \bibinfo {author} {\bibfnamefont {J.}~\bibnamefont {Collins}}, \bibinfo {author} {\bibfnamefont {C.}~\bibnamefont {Kaufhold}}, \ and\ \bibinfo {author} {\bibfnamefont {L.}~\bibnamefont {Theussl}},\ }\href {\doibase 10.1016/j.cpc.2009.02.020} {\bibfield  {journal} {\bibinfo  {journal} {Comput. Phys. Commun.}\ }\textbf {\bibinfo {volume} {180}},\ \bibinfo {pages} {1709} (\bibinfo {year} {2009})},\ \Eprint {http://arxiv.org/abs/0811.4113} {arXiv:0811.4113 [hep-ph]} \BibitemShut {NoStop}%
\bibitem [{\citenamefont {Rohatgi}(2022)}]{Rohatgi2022}%
  \BibitemOpen
  \bibfield  {author} {\bibinfo {author} {\bibfnamefont {A.}~\bibnamefont {Rohatgi}},\ }\href {https://automeris.io/WebPlotDigitizer} {\enquote {\bibinfo {title} {Webplotdigitizer: Version 4.6},}\ } (\bibinfo {year} {2022})\BibitemShut {NoStop}%
\end{thebibliography}%

\end{document}